\documentclass[a4paper,fleqn]{cas-sc}
\usepackage[numbers,sort&compress]{natbib}
\usepackage[utf8]{inputenc}
\usepackage[T1]{fontenc}
\usepackage{amsmath,amssymb}
\usepackage{mathtools}
\usepackage{graphicx}
\usepackage{booktabs}
\usepackage{multirow}
\usepackage{hyperref}
\usepackage{xurl}

\usepackage{textcomp}
\usepackage{lineno}
\usepackage{array}
\usepackage{xcolor}
\hypersetup{hidelinks}
\usepackage{tabularx}
\usepackage{makecell}
\usepackage{adjustbox}
\usepackage{microtype}
\usepackage{multicol}
\newcolumntype{L}[1]{>{\raggedright\arraybackslash}p{#1}}
\newcolumntype{C}[1]{>{\centering\arraybackslash}p{#1}}
\newcolumntype{Y}{>{\raggedright\arraybackslash}X}
\newcolumntype{Z}{>{\centering\arraybackslash}X}
\newcolumntype{R}{>{\raggedleft\arraybackslash}X}
\newcommand{\runinhead}[1]{\par\smallskip\noindent\textbf{#1.}\quad}

\newlength{\nmbox}
\newcommand{\nmhead}[1]{\par\addvspace{7pt}\noindent\textbf{#1}\par\addvspace{3pt}}
\newcommand{\nm}[2]{%
  \par\addvspace{2.6pt}\noindent
  \setbox0=\hbox{#1}%
  \hangindent=\dimexpr\nmbox+0.3em\relax\hangafter=1
  \ifdim\wd0>\nmbox #1\hspace{0.4em}#2%
  \else\hbox to \nmbox{#1\hfil}#2\fi\par}

\newcommand{\sitocline}[2]{\noindent\hyperref[#2]{#1}~\nobreak\dotfill~\pageref{#2}\par\addvspace{2pt}}

\graphicspath{{Figures/}}

\begin{document}
% Line numbers on for the initial submission / review copy.
%\linenumbers

\let\WriteBookmarks\relax
\def\floatpagepagefraction{1}
\def\textpagefraction{.001}

% Float placement. The multi-panel results figures render at ~0.65--0.87 of the
% text height at full width, which exceeds the default 0.7 top-float limit, so a
% top-only [pos=t] specifier cannot place them and they defer to the document
% end. Relaxing the fractions and allowing a dedicated float page (the figures
% use [pos=tp]) keeps each figure near its in-text reference.
\renewcommand{\topfraction}{0.92}
\renewcommand{\bottomfraction}{0.70}
\renewcommand{\textfraction}{0.06}
\renewcommand{\floatpagefraction}{0.55}
\setcounter{topnumber}{2}
\setcounter{bottomnumber}{1}
\setcounter{totalnumber}{3}

\shorttitle{Reactor cogeneration for data centers: a boundary analysis}
\shortauthors{Li et~al.}

\title[mode=title]{Techno-Economic Boundary Analysis of Small Modular Reactor Cogeneration for Hyperscale Data Center IT and Cooling Loads}

\author[1,3]{Honglin Li}[orcid=0000-0002-8405-3032]
% \ead{honglin.li@utdallas.edu}

\author[3]{Buxin She}
%\cormark[1]
% \ead{bshe@ksu.edu}

\author[1,2]{Jie Zhang}
\cormark[1]
% \ead{jiezhang@utdallas.edu}
% NOTE: the \ead lines above are commented out because cas-common.sty L386
% hard-codes \includegraphics{thumbnails/cas-email.jpeg}, which is not bundled
% with this template. Author emails are supplied through the journal submission
% system. To re-enable, place thumbnails/cas-email.jpeg (from the official CAS
% bundle) in the project directory and uncomment the lines.

\affiliation[1]{organization={Department of Mechanical Engineering, The University of Texas at Dallas},
                city={Richardson},
                postcode={75080},
                state={Texas},
                country={United States}}
\affiliation[2]{organization={Department of Electrical and Computer Engineering, The University of Texas at Dallas},
                city={Richardson},
                postcode={75080},
                state={Texas},
                country={United States}}
\affiliation[3]{organization={Department of Electrical and Computer Engineering, Kansas State University},
                city={Manhattan},
                postcode={66506},
                state={Kansas},
                country={United States}}

\cortext[cor1]{Co-corresponding author: J. Zhang (jiezhang@utdallas.edu).}

\begin{abstract}
Growing data-center electricity demand has increased interest in colocation with small modular reactors, but the conditions under which reactor power and steam-driven cooling reduce energy costs remain uncertain. Cooling can account for 20--40\% of facility electricity use, making steam extraction a potential source of savings as well as a loss of turbine generation. We co-optimize hourly reactor dispatch, steam extraction, absorption cooling and grid exchange for a 200~MW$_\mathrm{e}$ data center in the Electric Reliability Council of Texas (ERCOT) region, using a 109-entry scenario set spanning capital, market, policy, financing and cooling efficiency.
At 2023 mid-range reactor capital, the nuclear configurations cost 54.9--67.7\% more than grid supply even with the Section~45Y production tax credit. With the other 2023 assumptions held fixed, cogeneration reaches grid parity near \$5{,}000~kW$_\mathrm{e}^{-1}$. At nth-of-a-kind capital, nuclear supply is 85.9--98.7\% cheaper than grid procurement in 2023, but between parity and 34.8\% more expensive in the low-price 2024 market. A carbon price of \$53.1--64.0~tCO$_2^{-1}$ closes the mid-range gap when exports receive credits at the hourly average grid emission factor; the resulting negative balances are attributional and do not represent physical carbon removal. Absorption cooling responds to hourly electricity prices and supplies 32.6\% of annual cooling, but adds \$9.7~million~yr$^{-1}$ relative to the reactor-only plant. Absorption breaks even at an installed cost of \$26~kW$_\mathrm{c}^{-1}$ at baseline efficiency and \$529~kW$_\mathrm{c}^{-1}$ on a legacy-efficiency campus, compared with surveyed commercial prices of \$450--1{,}200~kW$_\mathrm{c}^{-1}$. Within the tested ranges, reactor capital and market conditions have the largest effects on competitiveness, while the value of heat recovery depends on absorber cost and the efficiency of the electric cooling it replaces.
\end{abstract}

\begin{highlights}
\item Reactor power and data-center cooling are co-optimized for a Texas campus.
\item Reactor options cost 54.9--67.7\% more than 2023 grid supply at mid-range capital.
\item At nth-of-a-kind capital, reactor options cost 85.9--98.7\% less than 2023 grid supply.
\item Carbon prices of \$53.1--64.0 per tonne bring reactors to parity under export crediting.
\item Absorption supplies 32.6\% of cooling; it undercuts reactor-only below \$26~kW$_\mathrm{c}^{-1}$.
\end{highlights}


\maketitle

\nolinenumbers
\begingroup
\small
\setlength{\columnsep}{2em}
\noindent{\large\textbf{Nomenclature}}\label{sec:nomen}\par\medskip
\begin{multicols}{2}
\raggedright

\nmhead{Sets and indices}
\nm{$t \in \mathcal{T}$}{hourly time step over an annual horizon, $|\mathcal{T}|=8760$}
\nm{$c \in \mathcal{C}$}{system configuration, $\mathcal{C}=\{0,1,2,3\}$}
\nm{$\mathcal{T}_\mathrm{out}$}{scheduled reactor-outage hours (baseline: 701~h from 15~March)}
\nm{$\mathcal{T}_\mathrm{maint}$}{scheduled absorption-chiller maintenance hours}

\nmhead{Decision variables (hour $t$)}
\nm{$P_\mathrm{rx}(t)$}{reactor thermal output (MW$_\mathrm{th}$)}
\nm{$P_\mathrm{tg}(t)$}{turbine gross electric output (MW$_\mathrm{e}$)}
\nm{$P_\mathrm{tn}(t)$}{turbine net electric output (MW$_\mathrm{e}$)}
\nm{$Q_\mathrm{ext}(t)$}{mid-pressure steam extraction to absorber (MW$_\mathrm{th}$)}
\nm{$Q_\mathrm{a}(t)$}{absorption-chiller cooling delivered (MW$_\mathrm{c}$)}
\nm{$P_\mathrm{v}(t)$}{vapor-compression-chiller electric input (MW$_\mathrm{e}$)}
\nm{$Q_\mathrm{v}(t)$}{vapor-compression-chiller cooling delivered (MW$_\mathrm{c}$)}
\nm{$P_\mathrm{g}^{+}(t)$}{grid import (MW$_\mathrm{e}$)}
\nm{$P_\mathrm{g}^{-}(t)$}{grid export (MW$_\mathrm{e}$)}
\nm{$P_\mathrm{ng}(t)$}{natural gas combined-cycle electric output, Case~3 (MW$_\mathrm{e}$)}
\nm{$B^{+}(t)$}{battery-energy-storage charging power (MW$_\mathrm{e}$)}
\nm{$B^{-}(t)$}{battery-energy-storage discharging power (MW$_\mathrm{e}$)}
\nm{$E_\mathrm{B}(t)$}{battery-energy-storage state of charge (MWh)}
\nm{$y(t)$}{grid import--export interlock binary (1 = import)}

\nmhead{Parameters}
\nm{$P_\mathrm{IT}(t)$}{data-center information-technology load (MW$_\mathrm{e}$)}
\nm{$\pi_\mathrm{g}(t)$}{Electric Reliability Council of Texas day-ahead locational marginal price (\$/MWh$_\mathrm{e}$)}
\nm{$\theta(t)$}{Electric Reliability Council of Texas hourly average carbon intensity (g\,CO$_2$/kWh$_\mathrm{e}$)}
\nm{$T_\mathrm{wb}(t)$}{Houston wet-bulb temperature ($^\circ$C)}
\nm{$\eta_\mathrm{r}$}{turbine rated electric efficiency at the design point (--)}
\nm{$a_\mathrm{w},\ b_\mathrm{w}$}{turbine Willans-line slope (MW$_\mathrm{e}$/MW$_\mathrm{th}$) and no-load intercept (MW$_\mathrm{e}$)}
\nm{$\alpha_\mathrm{w}$}{Willans extraction penalty (MW$_\mathrm{e}$/MW$_\mathrm{th}$)}
\nm{$\xi_\mathrm{aux}$}{turbine auxiliary fraction (--)}
\nm{$\eta_\mathrm{ch}$}{chilled-water chain efficiency (--)}
\nm{$\mathrm{COP}_\mathrm{a}(T_\mathrm{wb})$}{absorption-chiller coefficient of performance (--)}
\nm{$\mathrm{COP}_\mathrm{a}^{\,0},\ \overline{\mathrm{COP}}_\mathrm{a}$}{absorption coefficient of performance at design wet-bulb and nameplate cap (--)}
\nm{$T_\mathrm{wb,0},\ \delta$}{absorption design wet-bulb ($^\circ$C) and derating slope ($^\circ$C$^{-1}$)}
\nm{$\Delta T_\mathrm{cw},\ T_\mathrm{cry}$}{cooling-tower approach (K) and crystallization threshold ($^\circ$C)}
\nm{$a(t)$}{absorption availability (1 operating; 0 otherwise)}
\nm{$\mathrm{COP}_\mathrm{v}$}{vapor-compression-chiller full-load coefficient of performance (--)}
\nm{$f_\mathrm{v}(\cdot)$}{vapor-compression-chiller part-load electricity map (MW$_\mathrm{e}$)}
\nm{$\beta_\mathrm{a}$}{absorber parasitic load intensity (kW$_\mathrm{e}$/kW$_\mathrm{c}$)}
\nm{$\eta_\mathrm{rt}$}{battery-energy-storage round-trip efficiency (--)}
\nm{$\rho_\mathrm{B}$}{battery-energy-storage hourly self-discharge (--)}
\nm{$\mathrm{Cap}_\mathrm{E}$}{battery energy capacity (MWh)}
\nm{$\overline{\mathrm{CF}}$}{reactor annual capacity factor (--)}
\nm{$\overline{P}_\mathrm{rx}$, $\underline{P}_\mathrm{rx}$}{reactor max / min thermal output (MW$_\mathrm{th}$)}
\nm{$\Delta P_\mathrm{rx}^{\max}$}{reactor ramp limit (MW$_\mathrm{th}$/h)}
\nm{$\overline{P}_\mathrm{pcc}$}{point-of-common-coupling capacity (MW$_\mathrm{e}$)}
\nm{$S_k,\ N_k$}{installed size (MW or MWh) and capital-recovery period (yr) of asset $k$}
\nm{$A(i,n)$}{annuity present-value factor at rate $i$ over $n$ years (--)}
\nm{$\mathrm{CAPEX}_k$}{overnight capital coefficient of asset $k$}
\nm{$\mathrm{FOM}_k$}{fixed operations-and-maintenance coefficient of asset $k$}
\nm{$v_x$}{variable operations-and-maintenance coefficient of stream $x$}
\nm{$\mathrm{CRF}_k$}{capital recovery factor of asset $k$ (--)}
\nm{$\pi_\mathrm{f},\ \pi_\mathrm{f}^\mathrm{e}$}{nuclear fuel cost on thermal and net-electric bases (\$/MWh)}
\nm{$\pi_\mathrm{ng}(t)$}{delivered natural-gas price (Henry Hub + basis; \$/MMBtu)}
\nm{$h_\mathrm{ng},\ \eta_\mathrm{hhv}$}{natural gas combined-cycle full-load higher-heating-value heat rate (MMBtu/MWh$_\mathrm{e}$) and efficiency (--)}
\nm{$\overline{P}_\mathrm{ng}$}{natural gas combined-cycle rated electric capacity (MW$_\mathrm{e}$)}
\nm{$\phi_\mathrm{ng}(\ell)$}{natural gas combined-cycle part-load heat-rate multiplier (--)}
\nm{$\bar{c},\ c_{45\mathrm{Y}}$}{statutory and window-levelized 45Y credit (\$/MWh$_\mathrm{e}$)}
\nm{$Q_\mathrm{cool}^\mathrm{dem}(t)$}{data-center cooling demand (MW$_\mathrm{c}$)}
\nm{$\ell(t)$}{natural gas combined-cycle electrical load fraction (--)}
\nm{$\pi_\mathrm{CO_2}$}{carbon price (\$/tCO$_2$)}
\nm{$\Delta t$}{hourly time step length (h)}
\nm{$\gamma_\mathrm{rx},\ \gamma_\mathrm{ng}$}{reactor and natural gas combined-cycle lifecycle emissions (g\,CO$_2$-eq/kWh$_\mathrm{e}$)}
\nm{$\Theta$}{kg-to-tonne unit conversion ($=10^{-3}$)}

\nm{$\Phi_\mathrm{fuel}$}{annual nuclear-fuel cost (\$/yr)}
\nm{$\Phi_\mathrm{grid}$}{annual grid purchases less export revenue (\$/yr)}
\nm{$\Phi_\mathrm{CO_2}$}{annual carbon-accounting charge or credit (\$/yr)}
\nm{$\Phi_\mathrm{ptc}$}{annual clean-electricity production tax credit (\$/yr)}
\nm{$\Phi_\mathrm{fuel}^\mathrm{ng},\ \Phi_\mathrm{CO_2}^\mathrm{ng}$}{annual NGCC fuel and carbon costs (\$/yr)}
\nm{$M_c$}{case-specific grid-cost margin against the stated Case~0 benchmark (--)}
\nm{$\gamma_\mathrm{rx}^\mathrm{bill}$}{nuclear emissions factor used for carbon billing (g\,CO$_2$-eq/kWh$_\mathrm{e}$)}
\nm{$\eta_\mathrm{el},\ U$}{annual net-electric efficiency and electric-plus-heat utilization (--)}

\nmhead{Abbreviations}
\nm{ASHRAE}{American Society of Heating, Refrigerating and Air-Conditioning Engineers}
\nm{ATB}{Annual Technology Baseline}
\nm{ATB-Mid}{mid-range (Moderate) reactor-capital case in the National Laboratory of the Rockies ATB}
\nm{BESS}{battery energy storage system}
\nm{BWRX-300}{GE-Hitachi 870~MW$_\mathrm{th}$ / 300~MW$_\mathrm{e}$ boiling-water small modular reactor}
\nm{CAPEX}{overnight capital expenditure}
\nm{COP}{coefficient of performance}
\nm{CRF}{capital recovery factor}
\nm{EIA}{U.S. Energy Information Administration}
\nm{EPBT}{energy payback time (yr)}
\nm{ERCOT}{Electric Reliability Council of Texas}
\nm{FOAK / NOAK}{first-of-a-kind / nth-of-a-kind reactor capital cost}
\nm{FOM}{fixed operations and maintenance}
\nm{HHV}{higher heating value}
\nm{IT}{information technology}
\nm{LCOC}{levelized cost of cooling (\$/MWh$_\mathrm{c}$)}
\nm{LiBr--H$_2$O}{lithium bromide--water absorption pair}
\nm{LMP}{locational marginal price}
\nm{$M$}{grid-cost margin, one minus case total annualized cost divided by matched grid-only total annualized cost (positive = cheaper than grid)}
\nm{NGCC}{natural gas combined cycle}
\nm{NLCS}{net levelized cost of information-technology supply (\$/MWh$_\mathrm{e}$)}
\nm{NLR}{National Laboratory of the Rockies}
\nm{O\&M}{operations and maintenance}
\nm{OPG}{Ontario Power Generation}
\nm{PCC}{point of common coupling}
\nm{PTC}{production tax credit (Section~45Y of the Inflation Reduction Act)}
\nm{PUE}{power usage effectiveness}
\nm{SMR}{small modular reactor}
\nm{SOS2}{special ordered set of type 2}
\nm{TAC}{total annualized cost (\$/yr)}
\nm{UNECE}{United Nations Economic Commission for Europe}
\nm{VCC}{vapor-compression chiller}
\nm{VOM}{variable operations and maintenance}
\nm{WACC}{weighted-average cost of capital}

\end{multicols}
\endgroup
%\linenumbers

\section{Introduction}
\label{sec:intro}

% \subsection{Background and motivation}
% \label{sec:intro:background}

Data centers are again driving electricity demand growth after a decade in which efficiency gains moderated their global energy use \citep{masanet2020datacenter}. Worldwide, data-center electricity use is projected to roughly double to 945~TWh by 2030 \citep{iea2024electricity,iea2025energyAi}; in the United States (U.S.) it nearly doubled between 2019 and 2023 and could reach 6.7--12\% of national electricity use by 2028 \citep{shehabi2024}. Large-load interconnection requests in the Electric Reliability Council of Texas (ERCOT) already exceed conventional planning margins \citep{ercot2025longterm,eia2026aeo}. In addition to their annual energy demand, hyperscale campuses require near-continuous power and high availability, with their heaviest cooling duty occurring in hot, humid hours \citep{greenGrid2012pue,ashrae2021thermal}.

Small modular reactors (SMRs) have therefore re-entered data-center procurement discussions as firm, low-carbon generators whose unit sizes match a hyperscale campus more closely than conventional gigawatt-scale plants do \citep{mignacca2020smrFinance,doe2024liftoff,atb2024nuclear,asuega2023}. Their economics, however, are unsettled. Prior techno-economic assessments find that SMRs may reduce construction risk relative to large reactors yet still cost more than grid supply absent a carbon constraint \citep{asuega2023}, and capital-cost estimates span a wide first-of-a-kind (FOAK) to nth-of-a-kind (NOAK) range whose lower end depends on learning, standardization, financing and project delivery \citep{inl2024metaanalysis,nea2020constructionCost,atb2024nuclear}. Policy also affects the comparison. An eligible new-build reactor with zero operational emissions can receive the technology-neutral Section~45Y clean-electricity production tax credit (PTC). Whether this credit makes nuclear supply competitive depends on the cost difference before the credit is applied.

Thermal integration could further reduce electricity demand for cooling. Data-center cooling can account for 20--40\% of facility energy use, and absorption chillers can convert heat into chilled water when the heat source is sufficiently hot and available \citep{ebrahimi2014review,yuan2023wasteHeatReview,amiri2021}. Reactor steam cycles can provide the driving temperatures required by double-effect lithium bromide (LiBr) absorption chillers more readily than server exhaust. The value of this integration, however, depends on chiller capital cost, the opportunity cost of extracted steam, wet-bulb-driven derating, and backup electric cooling during hours of crystallization risk \citep{srikhirin2001absorption,herold2016absorption,cui2025switchable}. Treating heat recovery as an automatic benefit may therefore overstate the value of nuclear cogeneration for an efficiently cooled campus.

Recent nuclear--data-center assessments have examined mixed SMR and distributed-energy portfolios, demand-growth equilibrium, and integrated power--thermal architectures \citep{debnath2025smrDER,you2025dynamicEquilibrium,bhowmik2026nuclearDC,zhang2026nuclearDC}. For a colocated reactor, the value of thermal integration depends on cooling operation, electricity and gas prices, reactor capital, production incentives and carbon accounting. Evaluating these factors jointly can identify when a colocated reactor costs less than grid procurement and when absorption cooling recovers its additional capital cost.

Here, we present a techno-economic boundary analysis of a 200~MW$_\mathrm{e}$ data center colocated with a BWRX-300 SMR and a double-effect absorption chiller driven by steam extracted at the high- to low-pressure turbine crossover. The nuclear configurations are formulated as a mixed-integer linear program that co-optimizes hourly power supply, cooling and grid exchange against 2022--2024 ERCOT prices, an hourly information technology (IT) load profile, Houston wet-bulb temperatures, and daily Henry Hub spot prices plus basis applied to hourly gas dispatch. We compare four configurations: grid supply, reactor supply without heat recovery, reactor supply with absorption cooling, and an on-site natural gas combined cycle (NGCC). A 109-entry scenario set spans data-center scale, cooling efficiency, market year, battery deployment, reactor and absorption-chiller capital costs and their interaction with the market year, carbon price, and weighted-average cost of capital (WACC).

Under 2023 market conditions, the cogeneration plant remains more expensive than grid supply at mid-range reactor capital, even with the Section~45Y credit. Grid parity at zero carbon price requires reactor capital near \$5{,}000~kW$_\mathrm{e}^{-1}$; carbon pricing provides a second route at \$53.1--64.0~tCO$_2^{-1}$. These carbon-price thresholds assume export credits at the hourly average grid emission factor. The resulting negative balances do not represent physical carbon removal or a modeled marginal grid response. Absorption cooling follows hourly electricity prices but does not independently close the baseline cost gap.

The remainder of this paper is organized as follows. Section~\ref{sec:methods} describes the system configurations, the optimization formulation, the absorption submodel, the data inputs and the scenario design. Section~\ref{sec:results} presents the baseline economics and the boundary-condition sensitivities. Section~\ref{sec:discussion} synthesizes the boundary, discusses the implications for site selection and procurement, and states the limitations. Section~\ref{sec:conclusion} concludes. Scenario definitions, secondary endpoints and input provenance are documented in the Supplementary Information.

\section{Methodology}
\label{sec:methods}

\subsection{System configurations}
\label{sec:methods:system}

\begin{figure}[pos=tp]
\centering
\includegraphics[width=\linewidth]{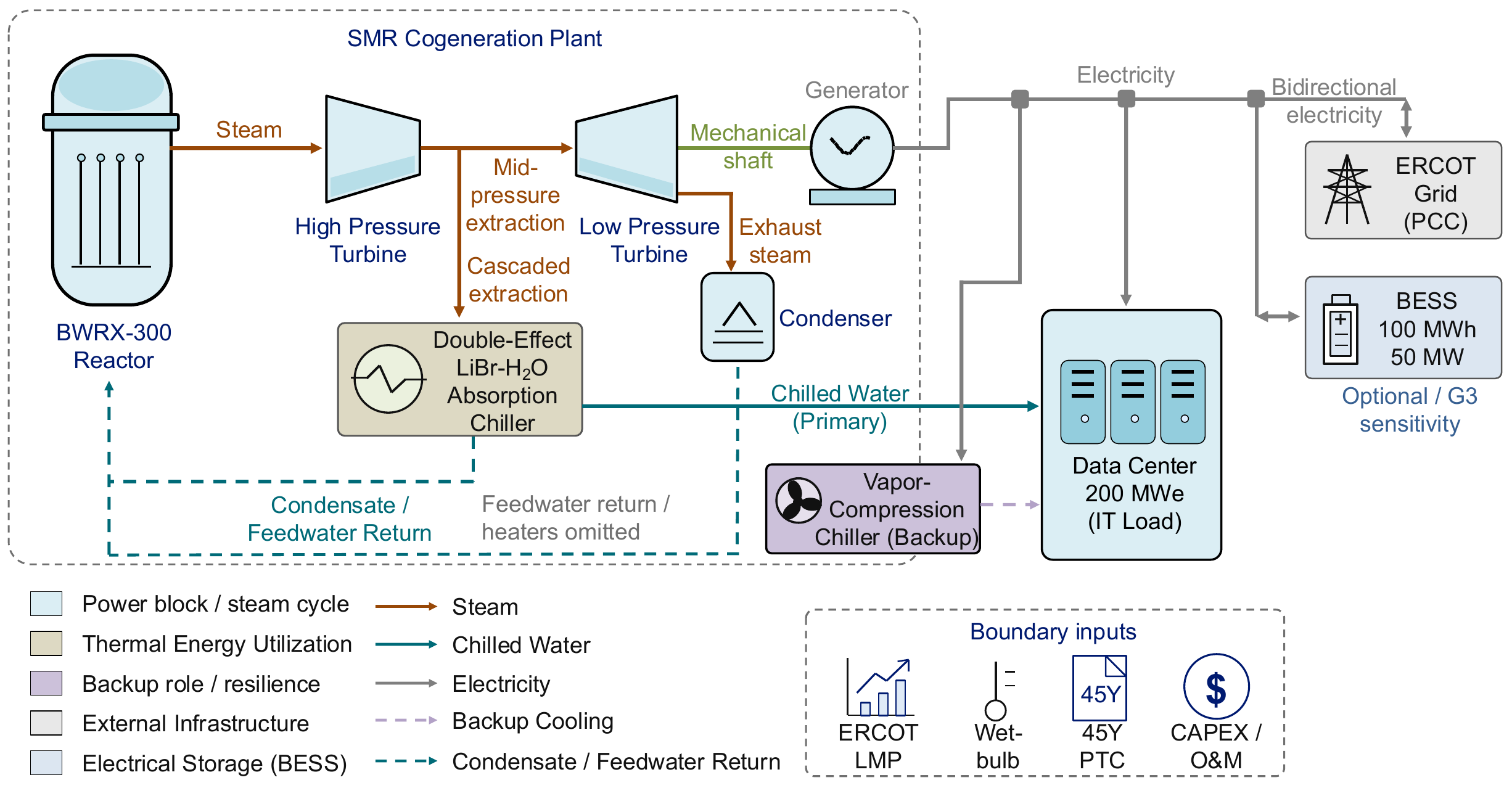}
\caption{Cascaded steam-extraction mechanism for SMR-powered data-center cooling. Steam from the BWRX-300 power block expands through the high-pressure turbine, a mid-pressure extraction stream drives a double-effect LiBr--H$_2$O absorption chiller, and the remaining steam expands through the low-pressure turbine to produce electricity. Chilled water supplies the data center, vapor-compression cooling shares the duty in low-price hours and covers the scheduled outage, and the grid and battery interfaces balance electricity. Complete case definitions, optimization equations and scenario groups are provided in the text and the Supplementary Information. PCC, point of common coupling; BESS, battery energy storage system; CAPEX, overnight capital expenditure; O\&M, operations and maintenance; LMP, locational marginal price; G3, battery sensitivity scenario group (Section~\ref{sec:methods:scenarios}).}
\label{fig:schematic}
\end{figure}

The analysis compares four configurations that supply electricity and chilled water to a single hyperscale data center in Houston, Texas, rated at 200~MW$_\mathrm{e}$ (Table~\ref{tab:cases}). All four share the same data-center load profile, vapor-compression chiller (VCC) technology where required and financial parameters, and differ only in the electricity supply pathway and in whether the cooling chain uses recovered reactor heat. Figure~\ref{fig:schematic} shows the fully integrated layout of Case~2; Cases~0, 1 and 3 are reduced versions in which selected blocks are disabled.

Case~0 represents the conventional grid-supplied data center. All electricity is purchased from ERCOT at the day-ahead locational marginal price (LMP), and all cooling is produced by a water-cooled centrifugal VCC with a design-point coefficient of performance (COP) of 3.2 at the Houston design wet-bulb. This case is the benchmark for the grid-cost margin reported throughout the paper. Case~1 replaces grid imports with on-site BWRX-300 generation; heat at the turbine condenser is rejected to the cooling tower and is not recovered. Bidirectional grid exchange is retained so that excess electricity can be sold to ERCOT and imports can cover demand during the scheduled reactor outage. Case~2 adds a 160~MW$_\mathrm{c}$ nominal double-effect LiBr--H$_2$O absorption unit fed by a mid-pressure extraction tap at the high-pressure to low-pressure (LP) crossover.
The cascade keeps the LP turbine in service, so diverted steam incurs the lost LP expansion work net of the hot-condensate-return credit.
A VCC of the same 160~MW$_\mathrm{c}$ capacity remains installed: it carries the full cooling load during the scheduled outage and whenever low electricity prices make steam diversion uneconomic, so the two chillers share dispatch rather than splitting into primary and emergency roles. Case~3 replaces the SMR with a 200~MW$_\mathrm{e}$ on-site F-class NGCC, retains the VCC and disconnects from the grid to isolate the value of replacing grid procurement with a fossil-fueled on-site alternative.

The grid-connected cases assume access to the stated PCC capacity; interconnection delays, site-specific connection charges and grid outages are outside the modeled operating boundary.

\begin{table}[pos=t]
\centering
\caption{System configurations modeled in this study. \checkmark{} denotes that the block is enabled; -- that it is disabled.}
\label{tab:cases}
\begin{tabular}{@{}llcccc@{}}
\toprule
 &  & Case~0 & Case~1 & Case~2 & Case~3 \\
\midrule
Generation & BWRX-300 SMR + steam turbine & --        & \checkmark & \checkmark & --        \\
           & NGCC on-site, F-class        & --        & --         & --         & \checkmark \\
           & ERCOT grid (PCC, 300 MW)     & \checkmark\,(import) & \checkmark\,(bidirectional) & \checkmark\,(bidirectional) & --        \\
Cooling    & VCC, water-cooled            & \checkmark\,(primary) & \checkmark\,(primary) & \checkmark\,(shared)  & \checkmark\,(primary) \\
           & Double-effect LiBr absorber  & --        & --         & \checkmark            & --        \\
Storage    & Lithium-ion BESS, 100 MWh / 50 MW & -- & sensitivity & sensitivity & -- \\
\bottomrule
\end{tabular}
\end{table}

The data-center IT load follows an annual hourly trace scaled to a 200~MW$_\mathrm{e}$ facility from a 10~MW colocation power profile generated by a year-long, one-minute discrete-event simulation \citep{vercellino2026aiworkload}, with a peak hourly load of 142.4~MW$_\mathrm{e}$, an annual average of 94.7~MW$_\mathrm{e}$, and a mean-to-peak utilization of 0.66.
The trace is circularly shifted by whole days so that day-of-week and hour-of-day align with each market year's calendar.
The same trace is used in every case and every main scenario; an additional constant-load sensitivity holds annual IT energy fixed to isolate load shape. We treat the full IT electrical draw as heat that the cooling system must reject, so the hourly cooling demand is
\begin{equation}
\label{eq:cool_demand}
Q_\mathrm{cool}^\mathrm{dem}(t) \;=\; \frac{P_\mathrm{IT}(t)}{\eta_\mathrm{ch}},
\end{equation}
where $\eta_\mathrm{ch}=0.90$ is the chilled-water distribution efficiency. Each scenario is labeled by its full-load power usage effectiveness (PUE) under vapor-compression cooling, $1+1/(\eta_\mathrm{ch}\,\mathrm{COP}_\mathrm{v})$, which is 1.35 at the baseline design-point $\mathrm{COP}_\mathrm{v}=3.2$. The dispatch-realized annual PUE, $1+\sum_t P_\mathrm{cool}(t)\Delta t/\sum_t P_\mathrm{IT}(t)\Delta t$, is lower: about 1.30 under vapor-compression cooling, where part-load operation and cooler wet-bulb hours lift the chiller's COP above its design value, and about 1.20 in Case~2, where steam-driven cooling additionally displaces chiller electricity in high-price hours. The cooling-efficiency sweep (Section~\ref{sec:results:pue}) varies $\mathrm{COP}_\mathrm{v}$ so that full-load PUE spans $\{1.10, 1.30, 1.50\}$, from best-practice to legacy hyperscale data-center operation \citep{greenGrid2012pue,ashrae2021thermal}; the corresponding design-point $\mathrm{COP}_\mathrm{v}$ values are 11.1, 3.7 and 2.2, so the PUE~1.10 anchor represents an economized or liquid-cooled plant rather than a mechanical chiller. Throughout this paper PUE counts cooling power only, excluding uninterruptible-power-supply, distribution and lighting overheads, so the reported values understate facility PUE as defined by the Green Grid. The steam driving the absorber is excluded from $P_\mathrm{cool}$. Case~2 PUE therefore measures the reduction in cooling electricity, not thermodynamic efficiency.

\subsection{Optimization formulation}
\label{sec:methods:opt}

The four cases are formulated as separate dispatch problems on the same 8760-hour horizon. Cases~0 and 3 admit no dispatch choice and reduce to closed-form annual cost evaluations: Case~0 meets every hour's load with grid purchases at the prevailing price, and the islanded Case~3 must follow load with its single on-site generator. In Cases~1 and 2, reactor ramping and, when storage is enabled, battery state of charge couple successive hours. These dispatch problems are mixed-integer linear programs with two sources of integer structure: the special ordered set of type 2 (SOS2) representation of the non-convex VCC part-load curve and the hourly grid import--export interlock of Eq.~\eqref{eq:grid_interlock}. No unit-commitment or battery charge--discharge interlock binaries are introduced. Battery operation is constrained by power limits, state of charge, round-trip losses and throughput costs; simultaneous charging and discharging is negligible in the evaluated storage scenarios. Cases~1 and 2 are implemented in Pyomo 6.9 and solved with Gurobi 13.0, retaining the barrier root-relaxation method, a relative gap target of $10^{-4}$ and a fixed seed. Independent annual solves use one solver thread each. Fixed-capital changes and duplicate ledger entries reuse the corresponding optimized dispatch; their costs and objective bounds are adjusted consistently. The same accounting structure is applied to all four cases.

\subsubsection{Reactor and turbine block (Cases~1 and 2)}
The BWRX-300 reactor is modeled as a continuous thermal source bounded between $\underline{P}_\mathrm{rx}=435$ and $\overline{P}_\mathrm{rx}=870~\mathrm{MW}_\mathrm{th}$ (a 50\% lower bound consistent with the stated daily load-following range), ramp-limited to $\Delta P_\mathrm{rx}^{\max}=261~\mathrm{MW}_\mathrm{th}~\mathrm{h}^{-1}$ (0.5\%~min$^{-1}$) and held at zero through a scheduled outage (Supplementary Note~9). The rated outputs are approximately 316~MW$_\mathrm{e}$ gross and 300~MW$_\mathrm{e}$ net, corresponding to gross and net thermal efficiencies of 36.3\% and 34.5\%, respectively \citep{gehitachi2024bwrx300,gehitachi2026bwrx300}. The baseline outage window $\mathcal{T}_\mathrm{out}$ comprises 701 contiguous hours beginning 15~March, so full output in every online hour realizes $\overline{\mathrm{CF}}=0.92$, close to the mature U.S.\ light-water-reactor fleet average \citep{eia2025capacityfactors}. This window is an aggregate availability proxy, not the manufacturer's refueling duration: the BWRX-300 design specifies 12--24-month cycles with 10--15-day refueling outages \citep{gehitachi2026bwrx300}. During the outage the data center buys grid power at the prevailing price, so backup energy is priced. A controlled two-year sensitivity separates an illustrative 14-day refueling component from the remaining annual downtime, changes only the refueling interval, and retains annual absorber maintenance (Supplementary Note~10).

The Rankine steam cycle is represented by a cascaded high-pressure to low-pressure turbine whose gross output follows a Willans line in reactor thermal input. The Willans curve is calibrated to the manufacturer\textquotesingle s gross rating while retaining a fixed normalized part-load shape; it is an engineering approximation rather than a manufacturer-validated turbine map. The no-load intercept accounts for the decline in part-load efficiency. A constant-efficiency model would overstate gross output at minimum load by about 8.3\%. With slope $a_\mathrm{w}\approx0.391~\mathrm{MW}_\mathrm{e}~\mathrm{MW}_\mathrm{th}^{-1}$, intercept $b_\mathrm{w}\approx24.2~\mathrm{MW}_\mathrm{e}$, and gross extraction penalty $\alpha_\mathrm{w}=0.20~\mathrm{MW}_\mathrm{e}~\mathrm{MW}_\mathrm{th}^{-1}$ for mid-pressure steam diverted to the absorber,
\begin{align}
P_\mathrm{tg}(t) \;=\; & a_\mathrm{w}\, P_\mathrm{rx}(t) \;-\; b_\mathrm{w} \;-\; \alpha_\mathrm{w}\, Q_\mathrm{ext}(t), \label{eq:turb_willans}\\
P_\mathrm{tn}(t) \;=\; & (1 - \xi_\mathrm{aux})\, P_\mathrm{tg}(t), \label{eq:turb_aux}
\end{align}
with $\xi_\mathrm{aux}=16/316\simeq0.0506$ and $P_\mathrm{tg}(t)=0$ during the outage window. The proportional auxiliary deduction is a design-point calibration to the gross and net ratings, not a measured off-design auxiliary-load curve. The extraction coefficient follows from a steam-table heat balance of the 7~bar / 165~$^\circ$C crossover tap: each kilogram of extracted steam forfeits about 480~kJ of low-pressure expansion work, delivers about 1{,}730~kJ to the absorber generator, and returns about 145~kJ of equivalent work through the hot-condensate credit to the feedwater train, giving $\alpha_\mathrm{w} \approx (480-145)/1730 \approx 0.19$, rounded to 0.20. Diverted steam still expands through the high-pressure turbine before extraction, so this is roughly half the cost of charging the full turbine efficiency against extracted heat, as would be appropriate for a parallel main-steam tap; Section~\ref{sec:results:cooling} reports the sensitivity to $\alpha_\mathrm{w}$ over 0.08--0.25.

\subsubsection{Absorption-chiller submodel (Case~2)}
The double-effect LiBr--H$_2$O absorption chiller is described by a time-varying COP and a binary availability indicator that captures LiBr crystallization risk \citep{srikhirin2001absorption,herold2016absorption}. With $T_\mathrm{wb,0}=26\,^\circ$C as the Houston design wet-bulb, $\mathrm{COP}_\mathrm{a}^{\,0}=1.10$ as the COP at the design wet-bulb, $\delta=0.015\,^\circ$C$^{-1}$ as the linear derating slope, and $\overline{\mathrm{COP}}_\mathrm{a}=1.30$ as the nameplate cap, the chiller COP at wet-bulb $T_\mathrm{wb}(t)$ is:
\begin{equation}
\label{eq:cop_abs}
\begin{aligned}
\mathrm{COP}_\mathrm{a}\!\big(T_\mathrm{wb}(t)\big) =
\max\Big\{
  \min\big[\mathrm{COP}_\mathrm{a}^{\,0}
  - \delta\,(T_\mathrm{wb}(t)-T_\mathrm{wb,0}), 
 \overline{\mathrm{COP}}_\mathrm{a}\big],\; 0.5\Big\}.
\end{aligned}
\end{equation}
The 0.5 lower bound is a numerical safeguard that never binds; the crystallization gate would engage first. With $\Delta T_\mathrm{cw}=5$~K as the cooling-tower approach and $T_\mathrm{cry}=34\,^\circ$C as the LiBr shutdown threshold on the cooling-water inlet (the 31~$^\circ$C Houston design point plus the datasheet's 3~K crystallization safety margin, inside the 34--35~$^\circ$C vendor envelope for double-effect machines), the availability indicator is:
\begin{equation}
\label{eq:abs_avail}
a(t) \;=\; \begin{cases} 1 & \text{if } T_\mathrm{wb}(t) + \Delta T_\mathrm{cw} \le T_\mathrm{cry} \text{ and } t \notin \mathcal{T}_\mathrm{out}\cup\mathcal{T}_\mathrm{maint}, \\ 0 & \text{otherwise}, \end{cases}
\end{equation}
so the absorption chiller is forced fully offline whenever the cooling-water inlet would exceed the crystallization threshold and throughout reactor outages and its own annual maintenance. The two windows coincide in the baseline; the non-refueling year of the two-year sensitivity retains the original absorber-maintenance window independently. With the $\Delta T_\mathrm{cw}=5$~K tower approach this becomes a wet-bulb gate at $T_\mathrm{cry}-\Delta T_\mathrm{cw}=29\,^\circ$C, the value marked in Figure~\ref{fig:inputs}a. Delivered absorption cooling is:
\begin{equation}
\label{eq:abs_yield}
Q_\mathrm{a}(t) \;=\; \mathrm{COP}_\mathrm{a}\!\big(T_\mathrm{wb}(t)\big)\cdot a(t)\cdot Q_\mathrm{ext}(t),
\end{equation}
subject to two explicit physical limits: the generator's steam-side capacity, sized to deliver nameplate cooling at the design-point COP ($Q_\mathrm{ext}(t) \le 160/1.10 \approx 145.5~\mathrm{MW}_\mathrm{th}$), and the installed cooling nameplate ($Q_\mathrm{a}(t) \le 160~\mathrm{MW}_\mathrm{c}$), which binds in cool hours when the COP rises above its design value. A parasitic electric load $\beta_\mathrm{a}\, Q_\mathrm{a}(t)$ with $\beta_\mathrm{a}=0.035~\mathrm{kW}_\mathrm{e}~\mathrm{kW}_\mathrm{c}^{-1}$ powers the solution and refrigerant pumps (0.020) and the cooling-tower fans (0.015). The VCC carries the residual cooling load.

Both cooling devices are represented by hourly quasi-steady operating relations. Wet-bulb and part-load effects describe off-design performance within each hour; chiller startup losses, thermal inertia and absorption-solution transients are not resolved. Reactor ramping and battery state of charge couple the overall dispatch across hours, but do not represent chiller transitions.

\subsubsection{Cooling, battery storage and grid blocks}
Vapor-compression cooling links its electric input to its cooling output through a part-load COP. A water-cooled centrifugal chiller gains efficiency at reduced load, with the COP peaking near 40--50\% of rated duty before declining at very low load, following the integrated part-load value of American National Standards Institute/Air-Conditioning, Heating, and Refrigeration Institute (ANSI/AHRI) Standard~550/590 \citep{ahri2020standard550}. The chiller electricity is therefore a non-convex, piecewise-linear function of the cooling delivered, $P_\mathrm{v}(t)=f_\mathrm{v}\!\big(Q_\mathrm{v}(t)\big)$, anchored at the design-point value $\mathrm{COP}_\mathrm{v}=3.2$ at the 26~$^\circ$C Houston design wet-bulb. The VCC uses the same relative wet-bulb response as the absorption chiller. Its design-point COP is multiplied each hour by $\min\!\big[1+0.0136\,(26-T_\mathrm{wb}(t)),\,1.18\big]$. The slope (0.015/1.10 per kelvin) and cap (1.30/1.10) are the absorber values normalized by its design-point COP, giving both chillers the same relative response to condenser-water temperature. This shared temperature response is a simplifying assumption; chiller-specific performance maps could change the estimated cooling savings. In the closed-form Cases~0 and 3 the chiller is evaluated directly at the realized load fraction; in the nuclear Cases~1 and 2 the same curve enters the optimization as an SOS2 piecewise map. The lithium-ion battery energy storage system (BESS), sized at 100~MWh and 50~MW (C/2), follows standard state-of-charge dynamics with round-trip efficiency $\eta_\mathrm{rt}=0.85$, hourly self-discharge $\rho_\mathrm{B}=4.17\times10^{-5}$\,h$^{-1}$, state-of-charge bounds $[0.05,\,0.95]\,\mathrm{Cap}_\mathrm{E}$ and a cyclically restored end state (Supplementary Note~9). Grid exchange is bounded by the point of common coupling (PCC) and interlocked by an hourly binary $y(t)$,
\begin{equation}
\label{eq:grid_interlock}
0 \le P_\mathrm{g}^{+}(t) \le \overline{P}_\mathrm{pcc}\, y(t), \qquad
0 \le P_\mathrm{g}^{-}(t) \le \overline{P}_\mathrm{pcc}\, \big(1 - y(t)\big),
\end{equation}
so the site never buys and sells in the same hour; simultaneous trade would be cost-neutral at a single hourly price and would leave gross trade volumes degenerate. $\overline{P}_\mathrm{pcc}=300$~MW for the baseline campus; its treatment in the size-matching sweep is described under scenario design.

\subsubsection{Energy balances}
Two equality constraints hold every hour. The electric balance equates supply, BESS discharge and grid imports to the IT load, chiller drives, absorber parasitics, BESS charge and grid exports:
\begin{equation}
\begin{aligned}
P_\mathrm{tn}(t) + P_\mathrm{g}^{+}(t) + B^{-}(t) \;=\;{}
& P_\mathrm{IT}(t) + P_\mathrm{v}(t) + \beta_\mathrm{a} Q_\mathrm{a}(t) \\
& {}+\; P_\mathrm{g}^{-}(t) + B^{+}(t).
\end{aligned}
\end{equation}
The cooling balance equates the two chillers' combined output, $Q_\mathrm{a}(t)+Q_\mathrm{v}(t)$, to the cooling demand of Eq.~\eqref{eq:cool_demand}. For Cases~0, 1, and 3 the absorption variables $Q_\mathrm{a}(t)$, $Q_\mathrm{ext}(t)$ are fixed to zero; for Case~3 the reactor and turbine variables are removed and the IT and chiller load is served entirely by the NGCC block at a full-load higher-heating-value (HHV) efficiency $\eta_\mathrm{hhv}=0.495$, with the part-load heat-rate multiplier $\phi_\mathrm{ng}$ introduced below.

\subsubsection{Objective and cost accounting}
The optimization minimizes total annualized cost (TAC), comprising overnight capital expenditure (CAPEX), fixed operations and maintenance (FOM), variable operations and maintenance (VOM), fuel, grid exchange, carbon and the production tax credit. Asset costs are annualized using a capital recovery factor (CRF):
\begin{equation}
\label{eq:tac}
\begin{aligned}
\min\; \mathrm{TAC} ={}&
\sum_k \mathrm{CRF}_k\,\mathrm{CAPEX}_k S_k + \mathrm{FOM} + \mathrm{VOM}\\
& + \Phi_\mathrm{fuel} + \Phi_\mathrm{grid} + \Phi_\mathrm{CO_2} - \Phi_\mathrm{ptc},
\end{aligned}
\end{equation}
where the first term annualizes CAPEX across reactor, turbine, absorption and VCC chillers, BESS and NGCC sized at $S_k$, each amortized over its capital-recovery period $N_k$ through an asset-specific CRF (reactor island 40~yr, chillers 25~yr, NGCC 30~yr, battery 15~yr). The reactor capital-recovery period is a financial assumption of this study, distinct from the 60-year technical life in the National Laboratory of the Rockies (NLR) Annual Technology Baseline (ATB) and the manufacturer\textquotesingle s 60-year design life with possible extension to 80 years \citep{atb2024nuclear,gehitachi2026bwrx300}. The 60- and 80-year capital-recovery sensitivities retain the same discount rate and PTC window (Supplementary Note~10); $\mathrm{FOM}=\sum_k \mathrm{FOM}_k S_k$ aggregates fixed costs, and VOM prices each stream's hourly throughput at its coefficient $v_x$ (Table~\ref{tab:equipment}). $\Phi_\mathrm{fuel}$ charges nuclear fuel on reactor thermal output at $\pi_\mathrm{f}$ (Supplementary Note~9), and the grid and carbon terms are:
\begin{align}
\Phi_\mathrm{grid} ={}&
  \sum_t \pi_\mathrm{g}(t)\big[P_\mathrm{g}^{+}(t) - P_\mathrm{g}^{-}(t)\big]\Delta t, \label{eq:grid_cost}\\
\Phi_\mathrm{CO_2} ={}&
  \pi_\mathrm{CO_2}\Theta \sum_t
  \Big[\gamma_\mathrm{rx}^\mathrm{bill} P_\mathrm{tn}(t) + \theta(t)\big(P_\mathrm{g}^{+}(t)-P_\mathrm{g}^{-}(t)\big)\Big]\Delta t. \label{eq:carbon_cost}
\end{align}
The nuclear fuel price $\pi_\mathrm{f}$ is charged on thermal energy because extraction does not reduce fuel burn, but it is anchored on the net-electric basis the fleet statistics use: $\pi_\mathrm{f} = \pi_\mathrm{f}^\mathrm{e}\,(\overline{P}_\mathrm{tn}/\overline{P}_\mathrm{rx})$ with $\pi_\mathrm{f}^\mathrm{e} = \$10~\mathrm{MWh}_\mathrm{e}^{-1}$ \citep{li2026nuclearH2}, i.e., \$3.45~MWh$_\mathrm{th}^{-1}$ at the plant's 34.5\% net efficiency. 

Case~3 uses the same cost components, with NGCC terms replacing the reactor terms. The combined-cycle heat rate rises below the design point, so fuel use and combustion emissions are both scaled by a part-load multiplier $\phi_\mathrm{ng}(\ell)\ge 1$ of the electrical load fraction $\ell(t)=P_\mathrm{ng}(t)/\overline{P}_\mathrm{ng}$, taken from a representative F-class combined-cycle characteristic extended below the 40\% gas-turbine stable-load point with the steeper low-load branch of published combined-cycle data, and equal to one at full load. The VOM term uses $v_\mathrm{ng}P_\mathrm{ng}(t)+v_\mathrm{v}Q_\mathrm{v}(t)$, where $P_\mathrm{ng}(t)$ is the NGCC electric output required by the IT and VCC loads, and the fuel and carbon terms convert electric output into gas purchases and combustion emissions through $h_\mathrm{ng}\,\phi_\mathrm{ng}(\ell)$ and $\gamma_\mathrm{ng}\,\phi_\mathrm{ng}(\ell)$, respectively (Supplementary Note~9), with full-load heat rate $h_\mathrm{ng}=3.412/\eta_\mathrm{hhv}$~MMBtu~MWh$_\mathrm{e}^{-1}$ and full-load lifecycle factor $\gamma_\mathrm{ng}=420$~g~CO$_2$-eq~kWh$_\mathrm{e}^{-1}$, equal to 360 direct-combustion and 60 upstream-methane g~CO$_2$-eq~kWh$_\mathrm{e}^{-1}$ \citep{alvarez2018methane}. Applying the same $\phi_\mathrm{ng}$ to both terms holds the fuel carbon content constant. 

The carbon term $\Phi_\mathrm{CO_2}$ prices the selected accounting balance by a uniform price $\pi_\mathrm{CO_2}$ and a unit-conversion factor $\Theta=10^{-3}$~t~CO$_2$~kg$^{-1}$. The baseline sets $\gamma_\mathrm{rx}^\mathrm{bill}=\gamma_\mathrm{rx}=12$~g~CO$_2$-eq~kWh$_\mathrm{e}^{-1}$, the generic nuclear lifecycle median reported by the IPCC, rather than a BWRX-300-specific measurement \citep{ipcc2014annexiii}. Nuclear direct operational CO$_2$ emissions are zero; pricing this lifecycle factor is a uniform accounting scenario, not a claim about an applicable tax. A separate sensitivity sets $\gamma_\mathrm{rx}^\mathrm{bill}=0$ while retaining $\gamma_\mathrm{rx}=12$ in the reported lifecycle balance (Supplementary Note~10). Exports credit the case at the ERCOT hourly average carbon intensity $\theta(t)$, derived from U.S. Energy Information Administration (EIA) Form 930 (EIA-930) generation-mix data \citep{eia930}, to reflect the time-varying emissions of the grid power exchanged, following time-resolved hourly emissions accounting rather than a static annual-average factor \citep{li2025ghgDemand}. This attributional accounting does not estimate the marginal grid-emissions response \citep{hawkes2010marginal,silerEvans2013regional}. 

The capital recovery factor $\mathrm{CRF}_k$ (Supplementary Note~9) is evaluated at the capital-recovery periods listed above and the baseline WACC $i=6.7\%$; the battery additionally carries a 5\% end-of-life salvage credit. The 6.7\% rate is the NLR ATB 2024 nominal financing assumption \citep{atb2024nuclear}, applied here to price and cost series held in constant dollars; this pairing overstates capital recovery relative to a fully real treatment (the implied real rate is about 4.4\% at the assumed 2.2\% inflation), so it is conservative for the capital-dominated nuclear cases. The WACC sensitivity (Section~\ref{sec:results:wacc}) sweeps $i$ over $\{5\%, 6.7\%, 10\%\}$.

\subsubsection{Clean-electricity production tax credit}
The nuclear cases earn the technology-neutral clean-electricity production tax credit (Section~45Y; 26~U.S.C.~\S\,45Y), enacted in the Inflation Reduction Act of 2022 \citep{ira2022}, on their generation \citep{usc26_45y,irs45y}: a new-build reactor placed in service after 2024 can qualify subject to statutory eligibility and timing conditions. At the prevailing-wage rate the inflation-adjusted credit for calendar 2025 is $\bar{c}=\$30~\mathrm{MWh}_\mathrm{e}^{-1}$ \citep{fr2025_45y}, flat in the market price with no gross-receipts phaseout, and payable for the statutory ten-year period from commissioning. The ten-year credit is levelized over a twenty-year reference window using the ratio of annuity present-value factors,
\begin{equation}
\label{eq:ptc}
c_{45\mathrm{Y}} \;=\; \bar{c}\,\frac{A(i,10)}{A(i,20)}, \qquad A(i,n)=\frac{1-(1+i)^{-n}}{i},
\end{equation}
which gives $c_{45\mathrm{Y}}=\$19.7~\mathrm{MWh}_\mathrm{e}^{-1}$ at the baseline $i=6.7\%$. The annual credit is $\Phi_\mathrm{ptc}=c_{45\mathrm{Y}}\sum_t P_\mathrm{tn}(t)\,\Delta t$, applied to net nuclear generation. We treat the credit as a baseline policy condition and hold the statutory amount at its calendar-2025 value. The credit is generation-indexed, so it raises the opportunity cost of every megawatt-hour the absorber diverts from the turbine, a coupling that shapes the cooling dispatch in Section~\ref{sec:results:cooling}. Cases~0 and 3 carry no nuclear generation and therefore no credit.

The cross-case figure of merit used throughout the rest of the paper is the grid-cost margin.
\begin{equation}
\label{eq:margin}
M_c \;=\; 1 \;-\; \frac{\mathrm{TAC}_c}{\mathrm{TAC}_0}\Big|_{\text{matched scenario}},
\end{equation}
defined as one minus the ratio of the case TAC to the Case~0 (grid-only) TAC at the same year and carbon price. In the cooling-efficiency sweep (group G1), where the swept chiller COP applies to the on-site cases, margins are quoted against the fixed full-load PUE-1.35 grid-only baseline---the same IT load served by a conventionally cooled facility. A positive margin means the configuration is cheaper than grid procurement; the magnitude is the relative cost difference. With the matched grid-only cost as denominator, margins from scenarios in which that cost itself changes---in particular across market years---are not directly comparable; cross-year results are therefore also characterized in absolute annualized cost.

\subsection{Data inputs}
\label{sec:methods:data}

\begin{figure}[pos=tp]
\centering
\includegraphics[width=\linewidth]{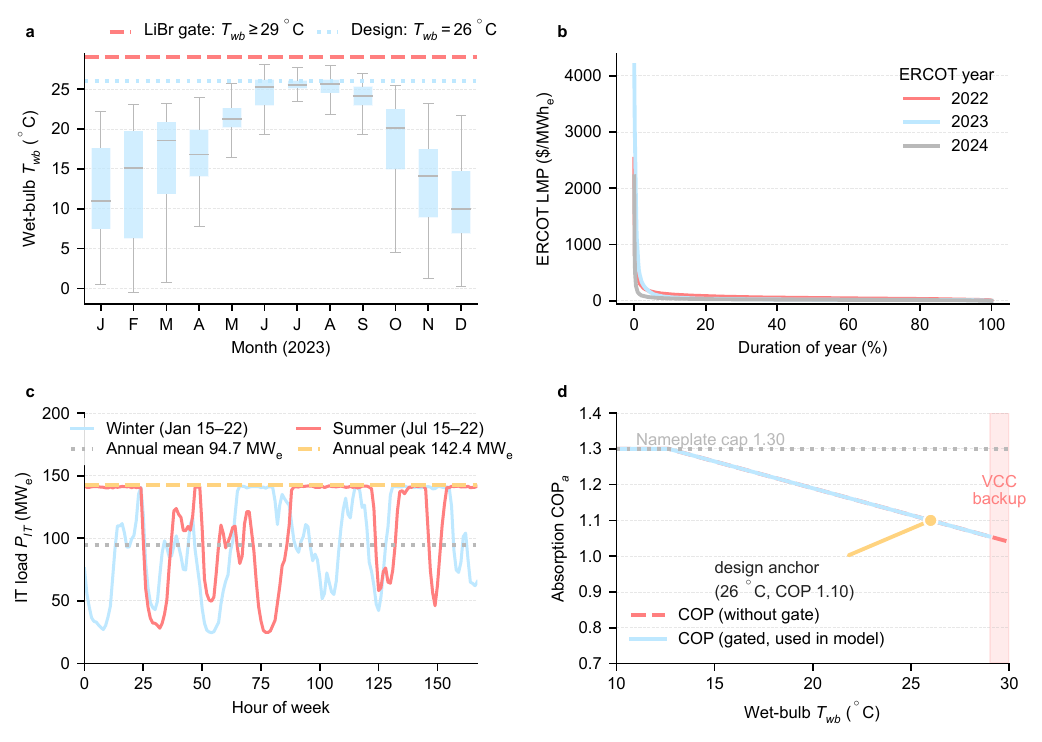}
\caption{Hourly input traces driving the optimization. (a)~Houston monthly wet-bulb distribution from the Open-Meteo historical-weather archive (year 2023). The dashed horizontal line marks the LiBr crystallization gate at $T_\mathrm{wb}\geq 29\,^\circ$C; the dotted line marks the COP design anchor $T_\mathrm{wb,0}=26\,^\circ$C, which the wet-bulb exceeds in about 32\% of summer hours. (b)~Duration curves of ERCOT day-ahead LMP for the three modeled years (2022--2024). (c)~Data-center IT-load profile for a typical winter and summer week, with annual mean and peak overlaid. (d)~Double-effect LiBr--H$_2$O absorption-chiller COP as a function of wet-bulb temperature; the solid curve is the gated COP used in the model, the dashed segment extends the linear derating past the 29~$^\circ$C crystallization gate, and the shaded band marks the backup-cooling region in which the VCC would carry the cooling load.}
\label{fig:inputs}
\end{figure}

Inputs to the optimization fall into three families: exogenous time series (hourly traces of IT load, electricity price, average carbon intensity, and wet-bulb temperature), equipment parameters (capital and operating cost coefficients plus technical bounds for each plant block) and financial parameters (project lifetime and weighted-average cost of capital). All runs use 8760-hour annual operating horizons. The four characteristic input traces are shown in Figure~\ref{fig:inputs}, the per-asset cost coefficients in Table~\ref{tab:equipment}, and the complete input inventory with source citations and the equipment technical parameters in Supplementary Tables~\ref{tab:data} and~\ref{tab:equipment_tech}.

The Houston wet-bulb temperature distribution (Figure~\ref{fig:inputs}a) \citep{openmeteo2023} drives the absorption-chiller availability constraint in Eq.~\eqref{eq:abs_avail}. Each market year is paired with its own wet-bulb, price and grid carbon-intensity records; the annual wet-bulb peak is 27.1~$^\circ$C in 2022 and 28.1~$^\circ$C in both 2023 and 2024, all below the 29~$^\circ$C LiBr crystallization gate, so the gate never engages in any of the three weather years. The ERCOT day-ahead LMP duration curves (Figure~\ref{fig:inputs}b), retrieved through gridstatus.io \citep{gridstatus}, capture the recent range of grid-procurement cost. We label the years by annual-mean LMP and gas conditions rather than by single-hour price spikes: 2022 is the high annual-mean gas / high annual-mean LMP regime, 2023 the baseline market year with moderate annual means, and 2024 the low annual-mean LMP regime. The hourly average carbon intensity $\theta(t)$, derived from the EIA-930 generation mix weighted by UNECE lifecycle factors \citep{eia930,unece2022lca}, has annual means of 320--350~g~CO$_2$~kWh$_\mathrm{e}^{-1}$ across the three years. The IT-load profile (Figure~\ref{fig:inputs}c) is a 10~MW NLR colocation workload trace scaled by 20$\times$ to the 200~MW facility \citep{vercellino2026aiworkload}, and the absorption-chiller COP curve (Figure~\ref{fig:inputs}d) plots Eq.~\eqref{eq:cop_abs} with its 29~$^\circ$C crystallization cutoff.

The per-asset cost coefficients carried through the CAPEX, FOM and VOM terms of Eq.~\eqref{eq:tac} are listed in Table~\ref{tab:equipment}. The three reactor capital anchors are \$14{,}700~kW$_\mathrm{e}^{-1}$ for FOAK, from the Ontario Power Generation (OPG) Darlington unit budget on an SMR-unit-only basis (shared four-unit site infrastructure excluded); \$7{,}615~kW$_\mathrm{e}^{-1}$ for the mid-range (Moderate) NLR ATB reactor-capital case (ATB-Mid), based on the NLR 2024 ATB value \citep{atb2024nuclear}; and \$2{,}250~kW$_\mathrm{e}^{-1}$ for NOAK, a GE-Hitachi long-run vendor target that we treat as a lower bound on plausible deployment cost rather than as a budgeted figure. The absorption-chiller capital cost grid is anchored at \$750~kW$_\mathrm{c}^{-1}$ baseline and brackets the literature range from \$450~kW$_\mathrm{c}^{-1}$ (Bare-Low, Chinese manufacturer ex-works) to \$1{,}200~kW$_\mathrm{c}^{-1}$ (Turnkey-High, Western turnkey engineering, procurement and construction). The NGCC cost block follows the NLR ATB 2024 F-class ATB-Mid (NLR Moderate) values; natural gas fuel cost is the U.S. EIA daily Henry Hub spot price \citep{eiaHenryHub}, forward-filled where needed and applied to each hourly dispatch interval, plus the Houston Ship Channel basis (mean \$0.10~MMBtu$^{-1}$). The data audit trail and provenance summary are provided in the Supplementary Information.

\begin{table}[pos=t]
\centering
\caption{Equipment cost and capital-recovery inputs used in the optimization. Each asset's CAPEX is annualized over its listed capital-recovery period through an asset-specific CRF; the battery additionally carries a 5\% end-of-life salvage credit. Reactor, NGCC and battery coefficients follow the NLR ATB 2024 \citep{atb2024nuclear}; the absorption range follows a U.S.\ Department of Energy combined-heat-and-power fact sheet \citep{doe2017chpAbsorption} and vendor literature; the vapor-compression value is a typical ASHRAE water-cooled centrifugal figure. Full provenance is given in Supplementary Note~6 and Supplementary Table~\ref{tab:data}.}
\label{tab:equipment}
\renewcommand{\arraystretch}{1.08}
\begin{tabularx}{\linewidth}{@{}L{0.26\linewidth}YL{0.17\linewidth}L{0.15\linewidth}c@{}}
\toprule
Asset & CAPEX & FOM & VOM & Period \\
\midrule
BWRX-300 reactor (Cases~1,~2) & \$7{,}615~kW$_\mathrm{e}^{-1}$ & \$121~kW$_\mathrm{e}^{-1}$~yr$^{-1}$ & \$3.50~MWh$_\mathrm{e}^{-1}$ & 40~yr \\
Cascaded steam turbine (Cases~1,~2) & bundled into reactor overnight capital cost & bundled & bundled & 40~yr \\
Double-effect absorption chiller (Case~2) & \$450--1{,}200~kW$_\mathrm{c}^{-1}$ (\$750 baseline) & \$20~kW$_\mathrm{c}^{-1}$~yr$^{-1}$ & \$0.40~MWh$_\mathrm{c}^{-1}$ & 25~yr \\
Vapor-compression chiller (Cases~0--3) & \$250~kW$_\mathrm{c}^{-1}$ & \$8~kW$_\mathrm{c}^{-1}$~yr$^{-1}$ & \$0.50~MWh$_\mathrm{c}^{-1}$ & 25~yr \\
Lithium-ion BESS (G3) & \$529~kWh$^{-1}$ (2-hour all-in) & \$26.5~kW$^{-1}$~yr$^{-1}$ & \$0.50~MWh$^{-1}$ & 15~yr \\
NGCC (Case~3) & \$1{,}330~kW$_\mathrm{e}^{-1}$ & \$29.3~kW$_\mathrm{e}^{-1}$~yr$^{-1}$ & \$2.65~MWh$_\mathrm{e}^{-1}$ & 30~yr \\
ERCOT interconnection (Cases~0--2) & not capitalized & not applicable & not applicable & not applicable \\
\bottomrule
\end{tabularx}
\end{table}

\subsection{Scenario design}
\label{sec:methods:scenarios}

Nine sensitivity groups are evaluated relative to the baseline 2023, full-load PUE~1.35, ATB-Mid CAPEX and WACC 6.7\% configuration to probe the boundary conditions of cogeneration economics. The scenario ledger, consisting of four baseline entries and 105 sensitivity entries, is given in Table~\ref{tab:scenarios}.

\begin{table}[pos=t]
\centering
\caption{Baseline and sensitivity scenario run ledger. C0--C3 denote Cases~0--3. Dashes (--) mark cases not included in that scenario group.}
\label{tab:scenarios}
\renewcommand{\arraystretch}{1.05}
\begin{tabularx}{\linewidth}{@{}llYrcccc@{}}
\toprule
Group & Sweep & Levels & Runs & C0 & C1 & C2 & C3 \\
\midrule
G0 & Baseline           & 2023, full-load PUE~1.35, ATB-Mid CAPEX, WACC 6.7\% & 4 & \checkmark & \checkmark & \checkmark & \checkmark \\
\midrule
G1 & Full-load PUE      & 1.10, 1.30, 1.50 ($\mathrm{COP}_\mathrm{v}=11.1$, 3.7, 2.2)          & 6   & -- & \checkmark & \checkmark & -- \\
G2 & ERCOT year         & 2022, 2023, 2024           & 12  & \checkmark & \checkmark & \checkmark & \checkmark \\
G3 & BESS               & on, off                    & 8   & control & \checkmark & \checkmark & control \\
G4 & SMR CAPEX          & FOAK, ATB-Mid, NOAK        & 6   & -- & \checkmark & \checkmark & -- \\
G5 & SMR $\times$ absorption CAPEX & 5 $\times$ 5 grid         & 25  & -- & -- & \checkmark & -- \\
G6 & Carbon price       & \$0, \$50, \$100~tCO$_2^{-1}$      & 12  & \checkmark & \checkmark & \checkmark & \checkmark \\
G7 & WACC               & 5, 6.7, 10\%               & 3   & -- & -- & \checkmark & -- \\
G8 & Data-center size   & 0.5--3.0$\times$ IT load   & 24  & \checkmark & \checkmark & \checkmark & \checkmark \\
G9 & Reactor capital $\times$ market year & NOAK, \$5{,}000~kW$_\mathrm{e}^{-1}$ $\times$ 2022, 2023, 2024 & 9 & -- & \checkmark & \checkmark & -- \\
\midrule
Total &                 &                            & 109 & & & & \\
\bottomrule
\end{tabularx}
\begin{flushleft}
\footnotesize \textit{Note:} In G3, ``control'' denotes duplicated control rows retained in the run ledger; BESS dispatch is optimized only for Cases~1 and~2 because Cases~0 and~3 are closed-form annual cost evaluations. In G9, the 2023 cells already solved in G4 and G5 are not double-counted, so the group comprises the eight off-2023 cells plus the Case~1 \$5{,}000~kW$_\mathrm{e}^{-1}$ 2023 anchor. The 109 entries are ledger rows rather than distinct optimization instances: rows that reproduce the baseline configuration inside a sweep---for example the 2023 cells of G2, the zero-carbon-price cells of G6 and the 1.0$\times$ cells of G8---are retained so that each group is complete; capital-only changes are evaluated from the same optimized dispatch by adjusting the fixed annuity. Additional robustness checks for load shape, outage scheduling, capital-recovery period and nuclear carbon billing are reported in Supplementary Note~10, separately from this 109-row ledger.
\end{flushleft}
\end{table}

The two-dimensional SMR $\times$ absorption capital grid (G5) in Table~\ref{tab:scenarios} identifies combinations for which Case~2 costs less than Case~0 at the 2023 baseline. The cooling-efficiency sweep (G1) changes the VCC full-load COP and reports the equivalent full-load PUE while the physical heat-rejection duty of Eq.~\eqref{eq:cool_demand} is held fixed, so it isolates the efficiency of the competing electric chiller rather than the size of the cooling load. The size sweep (G8) scales the IT trace and the data-center-side cooling and NGCC capacities from 0.5$\times$ to 3.0$\times$ while the commercial BWRX-300 unit stays fixed; the PCC is held at its 300~MW baseline below 1$\times$ and scales upward only above it, so unit-size matching is isolated without artificially constraining exports from a fixed reactor. The remaining groups sweep the market year, storage deployment, reactor capital, carbon price and WACC over the ranges listed in Table~\ref{tab:scenarios}.

\section{Results}
\label{sec:results}

The results compare each configuration with grid supply using the grid-cost margin of Eq.~\eqref{eq:margin}, then examine the cost of absorption cooling and the conditions for grid parity.

\subsection{Mid-range reactor capital keeps nuclear options above grid supply}
\label{sec:results:baseline}

\begin{figure}[pos=tp]
\centering
\includegraphics[width=\linewidth]{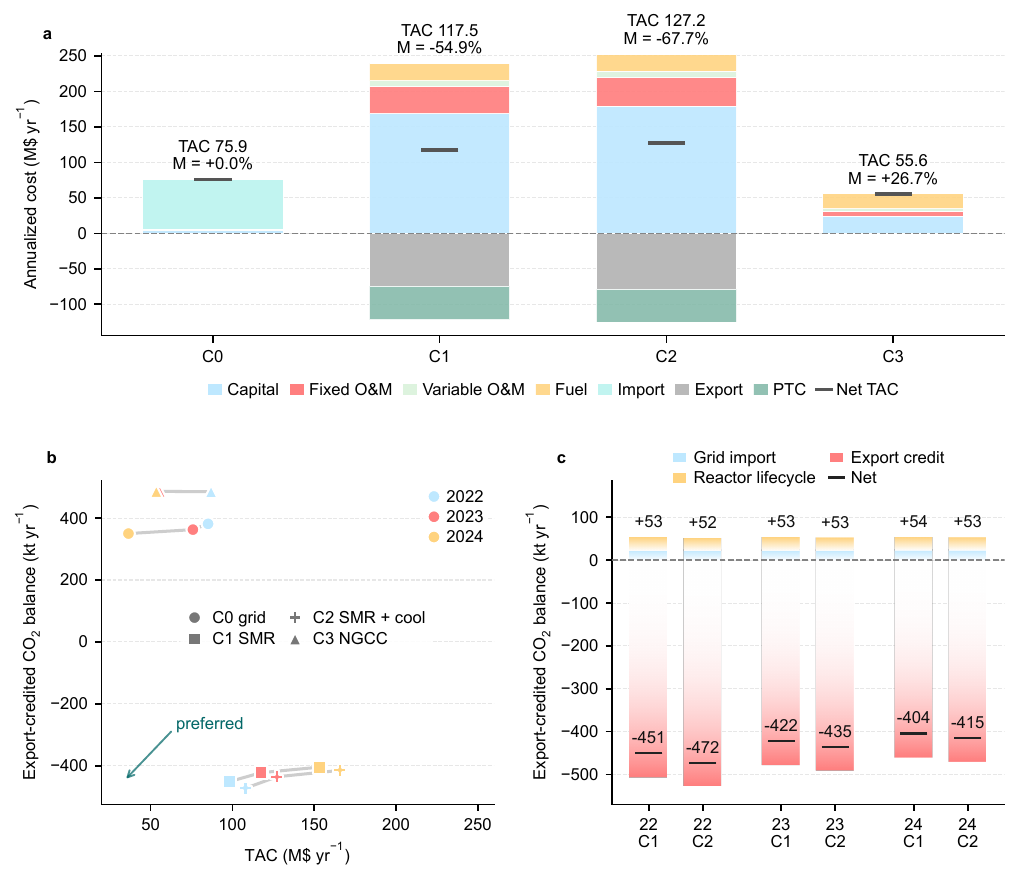}
\caption{Baseline economics of the four configurations under the 2023 ATB-Mid assumptions. The baseline uses a full-load PUE of 1.35 and a WACC of 6.7\%. (a)~TAC decomposition. Each bar stacks capital recovery, fixed and variable O\&M, fuel and grid imports. Bars below zero represent grid-export revenue and, for the nuclear cases, the Section~45Y PTC. The horizontal mark indicates net TAC, and $M$ is the grid-cost margin relative to Case~0. (b)~Cost--carbon trade-off across ERCOT years 2022, 2023 and 2024. Marker shape denotes the case, marker color denotes the year, and the arrow indicates the Pareto-preferred direction. (c)~Carbon-accounting components for the two nuclear configurations. Grid-import and reactor-lifecycle emissions are offset by grid-export credits, and the horizontal mark indicates annual export-credited CO$_2$ balance.}
\label{fig:baseline_economics}
\end{figure}

Under 2023 market conditions and the ATB-Mid reactor-capital assumption, the nuclear configurations are not at grid parity. The grid-supplied data center (Case~0) has a TAC of \$75.9~million~yr$^{-1}$, of which 93.3\% is the electricity bill. The reactor configurations without heat recovery (Case~1) and with absorption cooling (Case~2) have TACs of \$117.5 and \$127.2~million~yr$^{-1}$, respectively, even after receiving Section~45Y credits of \$47.6 and \$46.7~million~yr$^{-1}$ (Figure~\ref{fig:baseline_economics}a). Reactor capital recovery dominates both totals. Grid-export revenue and the PTC offset much of the fixed-cost burden, but the grid-cost margins remain $-54.9\%$ and $-67.7\%$.

The only baseline configuration that is cheaper than grid procurement is the on-site NGCC plant (Case~3), with a TAC of \$55.6~million~yr$^{-1}$ and a grid-cost margin of $+26.7\%$. This result reflects the 2023 fuel-price regime: delivered natural gas averaged \$2.63~MMBtu$^{-1}$, while ERCOT day-ahead prices were high enough for the grid-only electricity bill to exceed the cost of on-site generation. Case~3 is islanded and cannot import during the 41.9\% of hours when the day-ahead price falls below its full-load running-cost benchmark evaluated at the annual-mean fuel price. The comparison therefore reflects annual supply cost rather than hourly arbitrage. Adding absorption cooling to the nuclear plant increases TAC by \$9.7~million~yr$^{-1}$ relative to the reactor-only case.

Improving the match between reactor size and campus demand narrows the relative nuclear cost disadvantage but does not eliminate it. The modeled unit has a 300~MW$_\mathrm{e}$ net rating, about 3.2 times the data center's 94.7~MW$_\mathrm{e}$ average IT load, so Case~2 exports a net 1.37~TWh~yr$^{-1}$ to ERCOT. Scaling the campus brings the site to import--export balance between 2.0$\times$ and 2.5$\times$ and improves the matched Case~2 margin from $-67.7\%$ to $-31.1\%$ at 3.0$\times$, but it does not eliminate the ATB-Mid reactor-capital gap (Supplementary Note~1).

Figure~\ref{fig:baseline_economics}b,c shows the trade-off between annual cost and the export-credited carbon balance. Nuclear supply has lower carbon-accounting balances but higher costs than grid procurement in 2023 and 2024. Case~3 provides lower cost at the expense of higher emissions under 2023 gas prices. In Cases~1 and 2, export credits at the ERCOT hourly average carbon intensity exceed grid-import and reactor-lifecycle emissions. These credits give negative accounting balances, while both reactor configurations remain above grid cost at zero carbon price and mid-range reactor capital.

On an annual IT-energy basis, Case~2 has a net levelized cost of IT supply (NLCS) of \$153.3~MWh$_\mathrm{e}^{-1}$, compared with \$91.4~MWh$_\mathrm{e}^{-1}$ for grid procurement. Its energy payback time (EPBT) is 0.57~yr; water footprints and export-credited carbon balances are reported in Supplementary Note~2.

\subsection{Absorption cooling responds to electricity prices but adds no net value at the baseline}
\label{sec:results:cooling}

\begin{figure}[pos=tp]
\centering
\includegraphics[width=\linewidth]{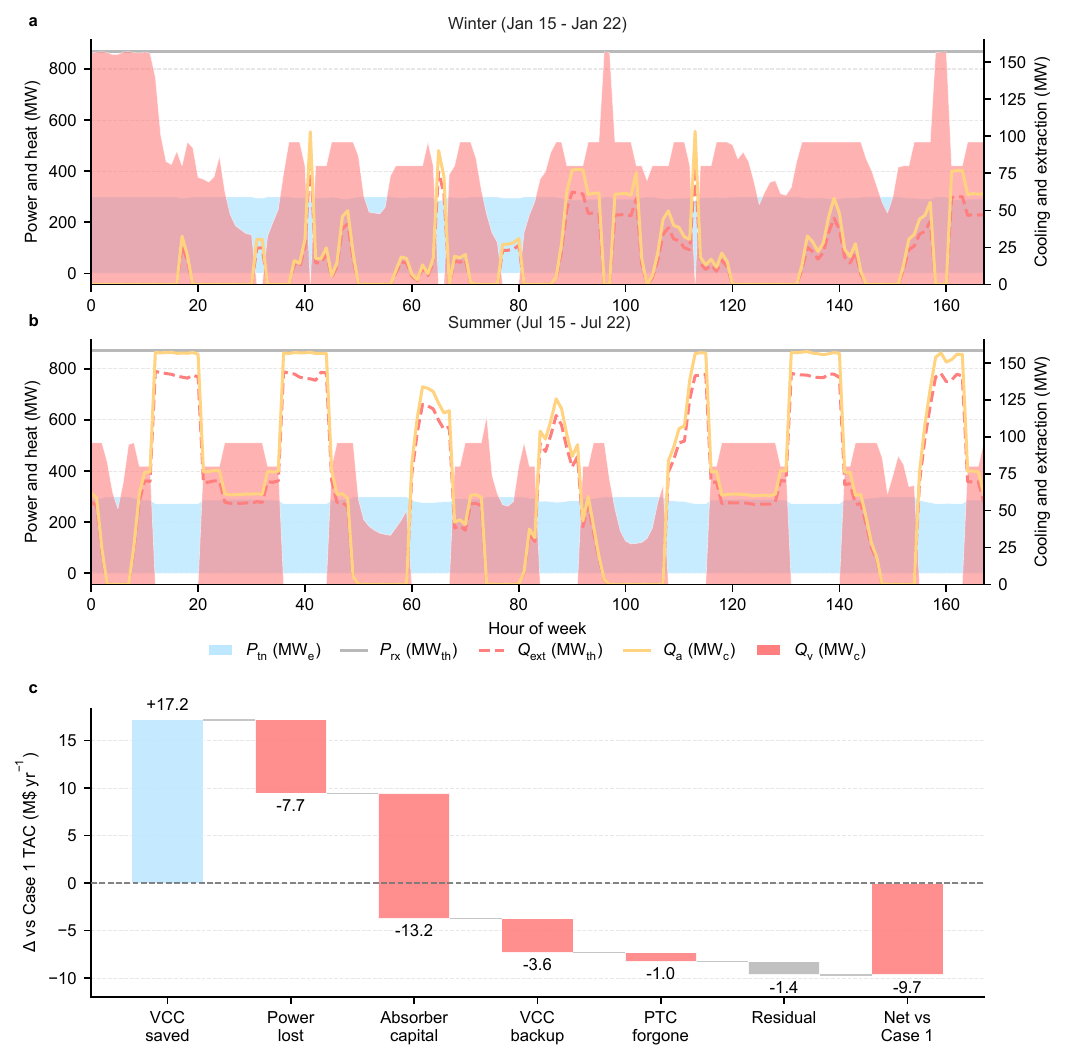}
\caption{Cogeneration operation and the marginal value of absorption cooling for Case~2. (a)~Hourly dispatch over a winter week (15--22 January) and (b)~a summer week (15--22 July). Reactor thermal output $P_\mathrm{rx}$ and turbine net electric output $P_\mathrm{tn}$ are read on the left axis; mid-pressure extraction $Q_\mathrm{ext}$, absorption cooling $Q_\mathrm{a}$ and VCC cooling $Q_\mathrm{v}$ are read on the right axis. The two chillers divide the load according to hourly electricity prices. Steam diversion forfeits the Section~45Y credit on displaced generation, so the VCC serves low-price winter hours and the absorber operates during high-price summer afternoons, subject to its wet-bulb-derated steam constraint. (c)~Waterfall decomposition of the marginal annual value of adding the absorption chiller to a reactor-only plant (Case~2 relative to Case~1). Positive bars reduce Case~2 TAC, whereas negative bars increase it. Retained VCC dispatch covers low-price hours and the scheduled outage. PTC forgone is the credit lost when steam diversion reduces turbine generation. The residual bar reconciles the component sum with the observed \$9.7~million~yr$^{-1}$ TAC difference. Bar labels are rounded independently, so they need not sum exactly to the net.}
\label{fig:operation_value}
\end{figure}

In the two representative weeks, the division of cooling between the two chillers responds to hourly electricity prices (Figure~\ref{fig:operation_value}a,b). Diverting steam to the absorber forfeits the Section~45Y credit on the turbine output it displaces, whereas operating the electric chiller incurs the prevailing LMP. The absorber is dispatched when the hourly price and the electric chiller's marginal electricity use at part load are high, while the VCC serves lower-price hours. During the January week, when the mean LMP is \$21.6~MWh$_\mathrm{e}^{-1}$, the absorber supplies only about one fifth of the cooling. During the July week, when afternoon prices exceed the credit-adjusted dispatch threshold, it supplies about three fifths. Wet-bulb derating can also limit absorption output in hot hours by lowering the COP and tightening the steam constraint. The reactor operates at nameplate thermal output throughout both plotted weeks, while the steam-side capacity and extraction coefficient limit the reduction in turbine output to about 27.6~MW$_\mathrm{e}$ from its 300~MW$_\mathrm{e}$ net rating.

Across the full year, the absorption chiller delivers 300.2~GWh$_\mathrm{c}$, or 32.6\% of annual cooling, over roughly 4{,}800 operating hours. The VCC delivers the remaining 621.6~GWh$_\mathrm{c}$, including the full cooling load during the 701-hour scheduled outage, when the absorber has no steam source and its own maintenance is scheduled. The crystallization gate of Eq.~\eqref{eq:abs_avail} never engages, in the baseline year or in either alternative market year (Section~\ref{sec:methods:data}). The extraction coefficient strongly changes the dispatch split: as $\alpha_\mathrm{w}$ increases from 0.08 to 0.25, the absorption share falls from 91.8\% to 11.7\%, while the Case~2 minus Case~1 cost gap increases from \$3.2 to \$11.5~million~yr$^{-1}$ (Table~\ref{tab:extraction}). The gap remains positive throughout this range at the baseline absorption capital cost.

\begin{table}[pos=t]
\centering
\caption{Extraction-penalty sensitivity at the 2023 ATB-Mid baseline. The coefficient is on a gross-electric basis; the net penalty is $(1-\xi_\mathrm{aux})\alpha_\mathrm{w}$.}
\label{tab:extraction}
\begin{tabular}{@{}lrrrr@{}}
\toprule
$\alpha_\mathrm{w}$ (MW$_\mathrm{e}$/MW$_\mathrm{th}$) & 0.08 & 0.15 & 0.20 & 0.25 \\
\midrule
Absorption share of cooling (\%) & 91.8 & 69.1 & 32.6 & 11.7 \\
Case~2 minus Case~1 TAC (\$M~yr$^{-1}$) & 3.2 & 7.3 & 9.7 & 11.5 \\
\bottomrule
\end{tabular}
\end{table}

Steam extraction raises electric-plus-heat utilization from 34.5\% in Case~1 to 37.5\% in Case~2, while net-electric efficiency falls from 34.5\% to 33.8\% (Supplementary Table~\ref{tab:secondary_kpis_cost_carbon}). Utilization counts net electricity and useful extracted heat at the reactor--steam-outlet boundary; downstream cooling is not counted again. Both cases have the same reactor thermal input and fuel cost, so the benefit is displaced electric cooling rather than reduced nuclear fuel consumption. The energy boundary and full annual quantities are given in Supplementary Note~10.

The waterfall in Figure~\ref{fig:operation_value}c decomposes the \$9.7~million~yr$^{-1}$ TAC difference between Cases~2 and 1 into the marginal components of adding absorption cooling to the reactor-only plant (Supplementary Table~\ref{tab:value_decomp}). The absorption chiller avoids \$17.2~million~yr$^{-1}$ of gross VCC electricity cost at the 2023 ERCOT hourly import price. These savings are offset by \$13.2~million~yr$^{-1}$ for absorption capital and fixed operations and maintenance, \$7.7~million~yr$^{-1}$ for the Willans-line electricity opportunity cost of extracted steam, \$3.6~million~yr$^{-1}$ for the VCC electricity retained during low-price hours and the scheduled outage, and \$1.0~million~yr$^{-1}$ of forgone PTC. A $-\$1.4$~million~yr$^{-1}$ residual from absorber auxiliary electricity and changes in variable operations and maintenance reconciles the component sum with the observed TAC difference. The monetized direct site-water increment is not included in the cost objective and is reported alongside the decomposition as a memo item of only \$0.01~million~yr$^{-1}$. The avoided VCC electricity cost does not cover the absorption chiller's capital cost and the opportunity cost of diverted steam.

\subsection{Lower electric-cooling efficiency reduces the absorption cost penalty}
\label{sec:results:pue}

The cooling-efficiency sweep (Supplementary Figure~\ref{fig:cooling_response}) varies the chiller COP so that the full-load PUE spans $\{1.10, 1.30, 1.50\}$ for Cases~1 and 2, with all other parameters held at the 2023 ATB-Mid baseline. At PUE~1.10, highly efficient electric cooling makes steam diversion uneconomic throughout the year, and the absorption chiller stays idle. The \$13.2~million~yr$^{-1}$ gap relative to Case~1 is therefore entirely attributable to the idle absorber's capital and fixed costs.

At PUE~1.30 and 1.50, the absorption chiller delivers 150.4 and 772.5~GWh$_\mathrm{c}$~yr$^{-1}$, respectively. As the displaced electric cooling becomes less efficient, the price threshold for economical steam diversion falls, and the absorber runs in more hours. Net cooling-electricity savings after the remaining VCC dispatch increase from zero at PUE~1.10 to \$10.2 and \$23.7~million~yr$^{-1}$ at PUE~1.30 and 1.50. Consequently, the Case~2 penalty relative to Case~1 narrows from \$13.2 to \$11.4 and \$3.0~million~yr$^{-1}$, but at baseline absorption capital it does not become a saving within the tested range. The Section~45Y credit contributes to this cost gap by increasing the opportunity cost of the turbine electricity displaced by steam extraction. Removing the credit from the existing cost decomposition narrows the PUE~1.50 gap to about \$0.51~million~yr$^{-1}$; this accounting comparison does not re-optimize dispatch without the credit.

Absorption capital enters the objective only as a dispatch-neutral annuity and Case~1 installs no absorber, so the gap between Cases~2 and~1 is exactly linear in the installed absorption cost, falling by \$13.4~million~yr$^{-1}$ per \$1{,}000~kW$_\mathrm{c}^{-1}$. Closing the baseline \$9.7~million~yr$^{-1}$ gap would therefore require an installed absorption cost near \$26~kW$_\mathrm{c}^{-1}$, about one tenth of the vapor-compression chiller capital coefficient. At full-load PUE~1.50 the requirement relaxes to about \$529~kW$_\mathrm{c}^{-1}$, inside the surveyed \$450--1{,}200~kW$_\mathrm{c}^{-1}$ range, where a Bare-Low absorber would save about \$1.1~million~yr$^{-1}$ against the reactor-only plant (Supplementary Note~3). Within the tested range, absorption becomes economical when low installed cost coincides with inefficient electric cooling. The absorber capacity is fixed at the campus cooling nameplate, so the analysis does not optimize its size as capital cost changes.

\subsection{Boundary sensitivities}
\label{sec:results:frontier}

\subsubsection{Market year}
\label{sec:results:marketyear}
The operating-year ERCOT price regime is the largest market sensitivity (Figure~\ref{fig:boundary_atlas}a and Supplementary Table~\ref{tab:market_years}). The 2022 results represent a high natural-gas-price regime. Delivered natural gas, including the Henry Hub price and Houston Ship Channel basis, averaged \$6.53~MMBtu$^{-1}$. With grid prices and on-site fuel costs rising together, the NGCC margin falls to $-2.2\%$, effectively reaching grid parity. The same high electricity prices increase the merchant value of nuclear exports. At mid-range reactor capital, the reactor-only Case~1 comes within 15.4\% of grid parity, its best market outcome in the solved envelope.

In 2023, low gas prices of \$2.63~MMBtu$^{-1}$ combined with firm ERCOT electricity prices make Case~3 the only on-site option that is cheaper than grid supply, with a margin of $+26.7\%$. By contrast, the 2024 results show the deepest negative margins for all on-site options. ERCOT LMPs fell to an annual mean of about \$28.3~MWh$_\mathrm{e}^{-1}$, making grid procurement comparatively inexpensive. The grid-only TAC falls to \$36.5~million~yr$^{-1}$, while the nuclear margins decline to $-320.2\%$ for Case~1 and $-354.5\%$ for Case~2. The nuclear margins move by 304.8--327.6 percentage points between the 2022 and 2024 regimes, but much of that swing is a denominator effect: the grid-only reference itself falls from \$85.1 to \$36.5~million~yr$^{-1}$. In absolute terms the market year moves the Case~2 total annualized cost by about \$57.6~million~yr$^{-1}$, roughly one-fifth of the \$270.5~million~yr$^{-1}$ spanned by the first-of-a-kind to nth-of-a-kind capital range.

\subsubsection{Reactor capital}
SMR overnight capital cost is the dominant technology-cost axis (Figure~\ref{fig:boundary_atlas}b). At FOAK capital of \$14{,}700~kW$_\mathrm{e}^{-1}$, the margins are $-257.8\%$ for Case~1 and $-270.5\%$ for Case~2. Reactor capital outweighs all other cost components, and the PTC only modestly narrows the gap. At ATB-Mid capital, the margins are $-54.9\%$ and $-67.7\%$, respectively. Even with the credit, the nuclear configurations cost about 1.5--1.7 times as much as grid supply. At NOAK capital of \$2{,}250~kW$_\mathrm{e}^{-1}$, the ranking reverses. The margins rise to $+98.7\%$ for Case~1 and $+85.9\%$ for Case~2. Adding the credit back to the optimized TAC still gives positive margins, with accounting pre-credit margins of $+35.9\%$ and $+24.4\%$. These capital-cost margins are solved under 2023 market conditions, and the low-capital result does not transfer unchanged across the market-year axis: evaluating the same capital anchors using the optimized dispatch for each of the other two regimes (scenario group G9) lifts the nth-of-a-kind margins to $+121.5\%$ and $+109.9\%$ in 2022 but drops them to $-0.5\%$ and $-34.8\%$ in 2024 (Supplementary Note~5). The nth-of-a-kind advantage is therefore conditional on the market regime, and the \$5{,}000~kW$_\mathrm{e}^{-1}$ parity threshold identified below is specific to the 2023 price year.

The \$47.6~million~yr$^{-1}$ credit received by Case~1 is worth about 62.8 percentage points of grid-cost margin at each capital level. It matters most in the middle of the capital-cost trajectory, where the pre-credit gap is comparable to the value of the credit. At ATB-Mid capital, it reduces the pre-credit Case~1 gap of $-117.7\%$ by about half and moves the parity threshold toward \$5{,}000~kW$_\mathrm{e}^{-1}$ (Figure~\ref{fig:boundary_atlas}d). At NOAK capital, Case~2 remains slightly less competitive than Case~1: absorption capital and the credit forgone on displaced generation still exceed the cooling-side savings. Nevertheless, both configurations remain well below grid cost.

\subsubsection{Financing}\label{sec:results:wacc}
\begin{figure}[pos=tp]
\centering
\includegraphics[width=\linewidth]{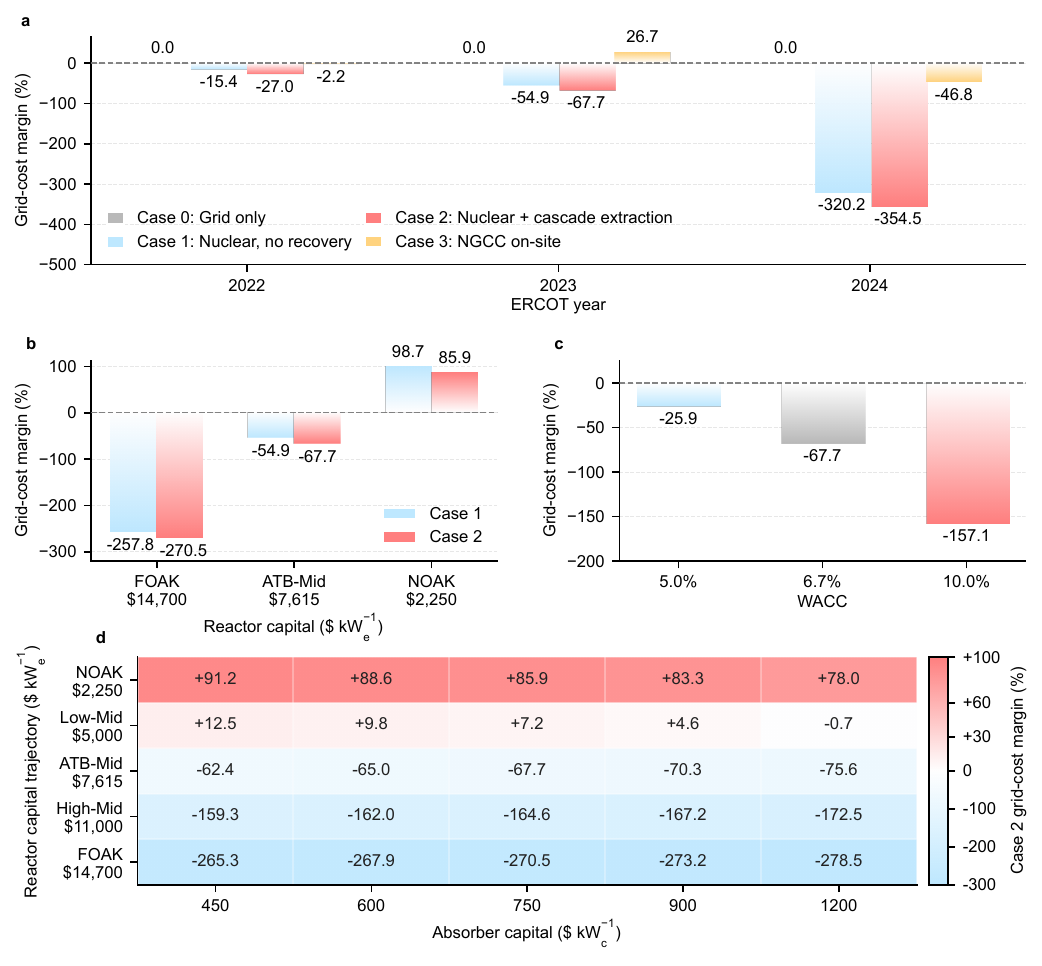}
\caption{Market, reactor-capital, financing and joint capital-cost boundary atlas. (a)~Grid-cost margin for the four configurations across ERCOT years 2022, 2023 and 2024. (b)~Grid-cost margin for Cases~1 and 2 across the FOAK, ATB-Mid and NOAK reactor-capital anchors: the OPG Darlington budget, NLR ATB 2024 Moderate value and GE-Hitachi vendor target, respectively. (c)~Case~2 grid-cost margin across the WACC sweep. (d)~Joint SMR and absorption-chiller capital-cost frontier for Case~2. Cell values report the grid-cost margin relative to the 2023 grid-only baseline. Bars or cells below zero indicate configurations that are more expensive than grid supply. Cell labels are rounded independently, so differences quoted in the text are computed from unrounded margins.}
\label{fig:boundary_atlas}
\end{figure}

Financing changes the annual capital charge and therefore the grid-cost margin (Figure~\ref{fig:boundary_atlas}c). Varying WACC from 5\% to 10\% for Case~2, with all other parameters fixed, raises TAC from \$95.6 to \$195.0~million~yr$^{-1}$ and moves the margin from $-25.9\%$ to $-157.1\%$. This 131.1-point span is smaller than the 356.5 points spanned by reactor capital, and Case~2 remains above grid cost throughout. Lower financing costs improve nuclear economics through the same annualized capital term as lower reactor construction costs.

\subsubsection{Joint capital-cost frontier}
The joint sensitivity in Figure~\ref{fig:boundary_atlas}d identifies combinations of reactor and absorption-chiller capital costs that bring Case~2 to grid parity. SMR capital cost spans the NOAK-to-FOAK range on the vertical axis (\$2{,}250--14{,}700~kW$_\mathrm{e}^{-1}$), while absorption-chiller capital cost spans the Bare-Low-to-Turnkey-High range on the horizontal axis (\$450--1{,}200~kW$_\mathrm{c}^{-1}$). All 25 cells in the $5\times5$ grid are evaluated under 2023 market conditions, a full-load PUE of 1.35, a WACC of 6.7\% and zero carbon price, with the PTC applied.

The competitive region includes the entire NOAK row and extends into the Low-Mid reactor row. Every cell in the NOAK row has a strongly positive grid-cost margin, ranging from $+91.2\%$ with Bare-Low absorption capital to $+78.0\%$ with Turnkey-High capital. The Low-Mid reactor row at \$5{,}000~kW$_\mathrm{e}^{-1}$ straddles grid parity, with margins ranging from $+12.5\%$ at Bare-Low absorption capital to $-0.7\%$ at Turnkey-High capital. The least favorable combination, a FOAK reactor with Turnkey-High absorption capital, has a margin of $-278.5\%$.

The frontier remains dominated by the reactor-cost axis. Across its full tested range, absorption capital changes the margin by no more than 13.2 percentage points. By contrast, moving from the \$5{,}000~kW$_\mathrm{e}^{-1}$ Low-Mid reactor row to the ATB-Mid row reduces the margin to between $-62.4\%$ and $-75.6\%$. A reactor capital cost of about \$5{,}000~kW$_\mathrm{e}^{-1}$ marks the grid-parity region for the cogeneration configuration under the Section~45Y credit, 2023 ERCOT prices and zero carbon price. No tested capital-cost combination at or above the ATB-Mid reactor cost reaches grid parity.

At baseline absorption capital, the continuous fixed-cost calculation gives reactor-capital parity at approximately \$5{,}250~kW$_\mathrm{e}^{-1}$ for Case~2 and \$5{,}700~kW$_\mathrm{e}^{-1}$ for Case~1.

\subsection{Carbon pricing closes the remaining mid-range gap}
\label{sec:results:policy}

\begin{figure}[pos=tp]
\centering
\includegraphics[width=\linewidth]{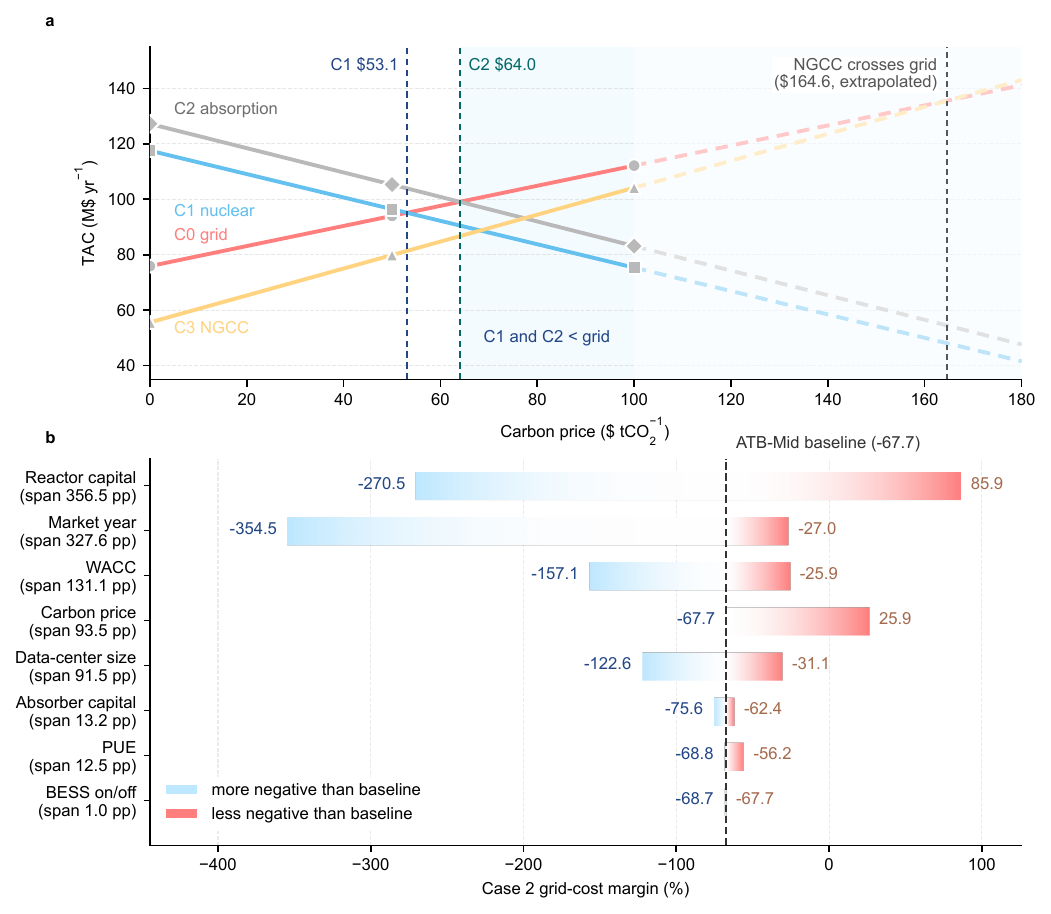}
\caption{Policy crossover and one-at-a-time sensitivity ranking. pp, percentage points. (a)~TAC as a function of carbon price for the four configurations at the 2023 ATB-Mid baseline, solid within the solved range of \$0--100~tCO$_2^{-1}$ and dashed where extrapolated. Vertical dashed lines mark where Cases~1, 2 and~3 cross the grid-only Case~0 baseline, at \$53.1, \$64.0 and \$164.6~tCO$_2^{-1}$; above the last, on-site NGCC is the most expensive of the four options. The shaded band, faint where extrapolated, spans the carbon prices at which both nuclear configurations are less expensive than grid supply. All crossovers credit merchant exports at the ERCOT hourly average emission factor. (b)~Ranked sensitivity of the Case~2 grid-cost margin to each boundary axis, anchored at the 2023 ATB-Mid baseline of $-67.7\%$. Bars extend from this baseline to the lower and upper margins on each axis. Because the margin is normalized by the matched grid-only cost, which itself falls across the market-year axis, the market-year bar is inflated relative to the others; the corresponding spans in absolute annualized cost are given in Section~\ref{sec:results:marketyear}. The data-center-size axis also rescales total demand. Bar-end labels are rounded independently, so the spans stated in the row labels are computed from unrounded margins.}
\label{fig:policy_sensitivity_summary}
\end{figure}

With reactor capital fixed at the ATB-Mid baseline, the policy sensitivity varies the carbon price over the range shown in Figure~\ref{fig:policy_sensitivity_summary}a. The annual export-credited CO$_2$ balances vary only modestly across the sweep; at zero carbon price they are: $+363$~kt~CO$_2$~yr$^{-1}$ for Case~0, $+486$~kt~CO$_2$~yr$^{-1}$ for Case~3, $-422$~kt~CO$_2$~yr$^{-1}$ for Case~1, and $-435$~kt~CO$_2$~yr$^{-1}$ for Case~2. Consequently, the relationship between TAC and carbon price is approximately linear for each configuration over the tested range. The PTC already keeps reactor generation near its maximum, leaving little additional export generation for a higher carbon price to induce.

At a fixed dispatch, a \$1~tCO$_2^{-1}$ increase in carbon price changes annual cost in proportion to the configuration's carbon-accounting balance. Grid-only and NGCC costs rise, while nuclear costs fall because merchant exports earn credits under the assumed hourly average grid attribution rule. The nuclear and NGCC trajectories cross the grid-only baseline in opposite directions as the carbon price rises. Case~1 reaches grid parity at approximately \$53.1~tCO$_2^{-1}$, while Case~2 reaches parity at approximately \$64.0~tCO$_2^{-1}$; both are linearly interpolated between solved carbon-price points within \$0--100~tCO$_2^{-1}$. At the upper bound of \$100~tCO$_2^{-1}$, Cases~1 and 2 are 32.8\% and 25.9\% less expensive than grid supply, respectively. Case~3 initially remains less expensive than grid supply but crosses above the grid-only baseline at an extrapolated carbon price of about \$164.6~tCO$_2^{-1}$, beyond the solved range. At NOAK reactor capital, both nuclear configurations are already less expensive than grid supply at a zero carbon price, a condition that keeps the carbon-pricing pathway distinct from the capital-cost one.

By replacing part of the electric cooling load with steam-driven cooling, Case~2 reaches a slightly more negative footprint than Case~1 ($-435$ versus $-422$~kt~CO$_2$~yr$^{-1}$), but its higher cost delays the grid-parity crossover from \$53.1 to \$64.0~tCO$_2^{-1}$. At baseline absorption capital, the more negative carbon balance does not offset the added cost of heat recovery within the tested carbon-price range.

Both crossover points rest on the crediting convention: the nuclear cases have negative accounting balances only because their roughly 1.34--1.40~TWh~yr$^{-1}$ of net exports are credited as displaced grid generation at the ERCOT hourly average intensity. Export-credit retention is not varied in the solved sweep (group G6), so the reported crossover prices are conditional on that convention.

Figure~\ref{fig:policy_sensitivity_summary}b ranks all eight boundary axes against the Case~2 ATB-Mid baseline margin of $-67.7\%$. Reactor capital and market year span 356.5 and 327.6 percentage points, WACC 131.1, and carbon price and data-center size 93.5 and 91.5; absorption-chiller capital, PUE and battery deployment span only 13.2, 12.5 and 1.0 percentage points, the last reflecting the battery's failure to recover its \$529~kWh$^{-1}$ capital cost anywhere in the sweep (Supplementary Note~4). The market-year bar is inflated by the denominator effect quantified in Section~\ref{sec:results:marketyear}; in absolute annualized cost, reactor capital remains the dominant axis.

\section{Discussion}
\label{sec:discussion}

% \subsection{Why capital decides}
% \label{sec:results:synthesis}

The sensitivity ranking in Figure~\ref{fig:policy_sensitivity_summary}b reflects the large fixed-cost contribution of reactor capital recovery and fixed O\&M. Hourly dispatch can reduce operating expenditure, but cannot change these fixed charges for an installed plant. Within the tested ranges, reactor capital and financing have larger effects on annual cost than cooling efficiency, absorption capital or battery deployment. The ranking depends on the ranges selected. The Section~45Y credit and carbon charges or credits depend on operating quantities. Near-baseload operation keeps annual generation and carbon balances relatively stable, but the payments vary with the credit rate and carbon price and affect the marginal value of generation and steam extraction.

% \subsection{Why heat recovery does not pay for itself}
% \label{sec:discussion:implications}
The higher electric-plus-heat utilization of cogeneration does not by itself establish an economic benefit. Extracted steam reduces electric cooling demand while sacrificing turbine output and the production credit earned on that output. At $\alpha_\mathrm{w}=0.20$~MW$_\mathrm{e}$~MW$_\mathrm{th}^{-1}$, a levelized credit of \$19.7~MWh$_\mathrm{e}^{-1}$ and the 2023 mean day-ahead price of \$57.3~MWh$_\mathrm{e}^{-1}$, each megawatt-hour of extracted heat has an electricity and credit opportunity cost near \$14.6~MWh$_\mathrm{th}^{-1}$, roughly a quarter of which comes from the credit. Generation-linked support therefore improves reactor economics while increasing the opportunity cost of steam-driven cooling.

The baseline decomposition assigns \$7.7~million~yr$^{-1}$ to forgone turbine electricity and \$1.0~million~yr$^{-1}$ to the associated PTC loss. Excluding these terms while holding dispatch fixed leaves a \$1.0~million~yr$^{-1}$ Case~2 cost penalty. Treating extracted steam as cost-free would therefore substantially overstate the value of absorption cooling. A different extraction design or support scheme could also change dispatch and would require a new optimization.

% \subsection{What transfers}
The reactor-to-campus size ratio also affects competitiveness. At the baseline, Case~2 exports a net 1.37~TWh~yr$^{-1}$, compared with annual IT demand of 0.83~TWh. Export revenue is therefore important even with substantial local consumption. Changes in the electricity market affect both the value of these exports and the cost of the grid-only alternative, explaining the strong dependence of the comparison on market year.

For heat recovery, the extraction coefficient, electric-chiller efficiency and absorber capital cost must be considered together. Lower extraction losses or less efficient electric cooling increase the operating benefit, but neither change covers the baseline absorber cost within the tested ranges. A lower-cost absorber can become economical when it replaces less efficient electric cooling. These results apply to the installed capacities evaluated here; joint optimization of reactor and chiller size could produce different cost thresholds.

Projected capital-cost reductions provide context for these thresholds. The 2024 ATB Moderate pathway assumes SMR capital-cost reductions of 19\% by 2035 and 34\% by 2040 relative to 2030, conditional on deployment-driven learning \citep{atb2024nuclear}. These technology-wide scenarios do not specify a BWRX-300 project cost or the year in which a particular site would become competitive. Policy support depends on construction-start timing and eligibility \citep{usc26_45y}, whereas the present calculations hold the calendar-2025 credit value fixed. The carbon-price threshold represents a conditional valuation, and future grid decarbonization would reduce the credit assigned to each exported megawatt-hour.

Grid-independent colocation may avoid connection delays and charges, but would forgo merchant-export revenue and require on-site supply for the full IT and cooling demand during refueling and forced outages. Diesel or other backup generators could provide this reserve, but their capacity, fuel, maintenance and availability costs would need to be included. Short-duration batteries cannot cover the energy requirement of a multi-day outage. Solar PV and hybrid wind--solar--storage portfolios offer other supply pathways, and renewables are expected to meet a substantial share of data-center demand growth \citep{iea2025energyAi}. Comparing these portfolios with reactor supply requires joint sizing of generation, storage and backup under the same hourly load, grid-access and reliability assumptions. The four configurations isolate the value of reactor supply and thermal integration without establishing a cost ranking against optimized renewable hybrids.

\subsection{Limitations}
\label{sec:results:limitations}
The results are subject to several modeling limitations. The IT trace is generated by a colocation-workload simulation rather than measured from the modeled campus. An equal-energy constant-load comparison leaves both nuclear configurations above grid cost, but does not establish applicability across data-center workloads (Supplementary Note~10). Hourly quasi-steady chiller maps omit startup losses and thermal inertia, so rapid switching and its operating savings may be overstated. The carbon balance uses hourly average rather than marginal grid emission factors with fixed market prices; its negative values are export-credit balances, not physical carbon removal or estimates of the emissions response to new generation. Demand charges, site-specific connection costs and battery ancillary-service revenues are excluded.

Reliability is not imposed on a common basis: nuclear cases include scheduled outages and grid purchases, whereas the gas comparator omits forced outages, redundancy and outage-covering grid access. None of the cost rankings establishes the cost of achieving 99.999\% availability, which requires consistent reserve and failure assumptions. Credit levelization and capital-recovery periods also affect the thresholds (Supplementary Notes~8 and~10). The Willans part-load curve and proportional auxiliary demand are engineering approximations; the stated nameplate calibration is not transient plant validation. Finally, a fixed single-unit reactor does not represent staged multi-module deployment against a growing campus load.

\section{Conclusion}
\label{sec:conclusion}

Within the tested ranges, reactor capital and market conditions have the largest effects on the grid-cost margin of colocated nuclear supply. The cogeneration configuration reaches grid parity near \$5{,}000~kW$_\mathrm{e}^{-1}$ of reactor capital under 2023 prices, baseline financing and the assumed Section~45Y credit. Cooling-efficiency, absorption-capital and battery sensitivities have smaller effects and do not remove the baseline nuclear cost disadvantage.

Absorption cooling increases electric-plus-heat utilization but adds annual cost at the baseline. The electricity savings do not cover the absorber's capital and fixed costs, forgone turbine output and the production credit lost on that output. Generation-linked support improves reactor economics while increasing the opportunity cost of heat recovery. The utilization gain does not reduce nuclear fuel use in the modeled dispatch.

These cost thresholds depend on market prices, financing and the treatment of policy credits. Two questions remain for further analysis: whether other thermal applications offer greater value for extracted steam, and how the preferred use of steam changes after generation-linked support expires.

\section*{CRediT authorship contribution statement}
\textbf{Honglin Li}: Conceptualization, Methodology, Software, Formal analysis, Investigation, Data curation, Visualization, Writing -- original draft, Writing -- review \& editing. \textbf{Buxin She}: Methodology, Validation, Writing -- review \& editing. \textbf{Jie Zhang}: Conceptualization, Supervision, Resources, Funding acquisition, Project administration, Writing -- review \& editing.

\section*{Declaration of competing interest}
The authors declare that they have no known competing financial interests or personal relationships that could have appeared to influence the work reported in this paper.

\section*{Data availability}
All data that support the findings of this study, including the input parameters, scenario definitions, run-level summaries, hourly dispatch traces and figure source data, are available in the project repository at \url{https://github.com/Henryutd2021/Nuclear-DC} and will be deposited in a persistent public archive (Zenodo) with a citable digital object identifier on acceptance. ERCOT day-ahead locational marginal prices are sourced from gridstatus.io under their public application programming interface terms; hourly average grid carbon intensities are derived from the EIA-930 generation mix and UNECE 2022 lifecycle emission factors; Houston wet-bulb temperature is sourced from the Open-Meteo historical-weather archive; reactor and NGCC capital costs follow the National Laboratory of the Rockies Annual Technology Baseline 2024; absorption-chiller capital costs are bracketed using published U.S. Department of Energy combined-heat-and-power fact-sheet data and publicly available vendor literature. No proprietary or restricted-access data are used in this study.

%\section*{Acknowledgements}
% Funding statement to be added at submission if required by the journal.

% Numeric house style for Elsevier energy journals. unsrtnat is natbib-native
% and universally installed; Elsevier reformats to the exact proof-stage layout
% at acceptance per "Your Paper Your Way".
\bibliographystyle{unsrtnat}
\bibliography{references}

\clearpage
% ===================== Supplementary Information =====================
% Counters renumbered with an S prefix; line numbers suppressed.
\setcounter{section}{0}
\renewcommand{\thesection}{S\arabic{section}}
\renewcommand{\thesubsection}{S\arabic{section}.\arabic{subsection}}
\renewcommand{\thefigure}{S\arabic{figure}}\setcounter{figure}{0}
\renewcommand{\thetable}{S\arabic{table}}\setcounter{table}{0}
\renewcommand{\theequation}{S\arabic{equation}}\setcounter{equation}{0}
\renewcommand{\theHsection}{S\arabic{section}}
\renewcommand{\theHsubsection}{S\arabic{section}.\arabic{subsection}}
\renewcommand{\theHfigure}{S\arabic{figure}}
\renewcommand{\theHtable}{S\arabic{table}}
\renewcommand{\theHequation}{S\arabic{equation}}
\nolinenumbers

\section*{Supplementary Information}

This Supplementary Information accompanies ``Techno-Economic Boundary Analysis of Small Modular Reactor Cogeneration for Hyperscale Data Center IT and Cooling Loads''. It contains ten Supplementary Notes, two Supplementary Figures and nine Supplementary Tables. Abbreviations are listed below; model variables and parameters are defined in the main article's Nomenclature.

\subsection*{Abbreviations}
{\small\noindent
ASHRAE, American Society of Heating, Refrigerating and Air-Conditioning Engineers;
ATB, Annual Technology Baseline;
ATB-Mid, mid-range (Moderate) reactor-capital case of the NLR Annual Technology Baseline;
BESS, battery energy storage system;
BWRX-300, GE-Hitachi 870~MW$_\mathrm{th}$ / 300~MW$_\mathrm{e}$ boiling-water small modular reactor;
COP, coefficient of performance;
EIA, U.S.\ Energy Information Administration;
EPBT, energy payback time;
ERCOT, Electric Reliability Council of Texas;
FOAK, first-of-a-kind;
G3 and G8, the battery and data-center-size sensitivity scenario groups defined in Table~\ref{tab:scenarios} of the main article;
HHV, higher heating value;
IT, information technology;
LCOC, levelized cost of cooling;
LMP, locational marginal price;
$M$, grid-cost margin;
\$M, million US dollars;
NGCC, natural gas combined cycle;
NLCS, net levelized cost of IT supply;
NLR, National Laboratory of the Rockies;
NOAK, nth-of-a-kind;
O\&M, operations and maintenance;
PCC, point of common coupling;
PTC, production tax credit (Section~45Y);
PUE, power usage effectiveness;
Section~45Y, the technology-neutral clean-electricity production tax credit of 26~U.S.C.~\S\,45Y;
SMR, small modular reactor;
TAC, total annualized cost;
UNECE, United Nations Economic Commission for Europe;
VCC, vapor-compression chiller;
WACC, weighted-average cost of capital.
\par}

\subsection*{Contents}
{\small
\sitocline{Supplementary Note 1: Data-center size-matching results}{note:size}
\sitocline{Supplementary Note 2: Secondary cost, carbon and water endpoints}{note:secondary}
\sitocline{Supplementary Note 3: Marginal value of absorption cooling}{note:absvalue}
\sitocline{Supplementary Note 4: Cogeneration dispatch and storage detail}{note:dispatch}
\sitocline{Supplementary Note 5: Market inputs by operating year}{note:market}
\sitocline{Supplementary Note 6: Exogenous data sources and provenance}{note:provenance}
\sitocline{Supplementary Note 7: Secondary endpoint definitions}{note:endpoints}
\sitocline{Supplementary Note 8: Production-tax-credit levelization robustness}{note:ptcwindow}
\sitocline{Supplementary Note 9: Model formulation detail and full input inventory}{note:formulation}

\sitocline{Supplementary Note 10: Energy utilization and operating-assumption sensitivities}{note:robustness}
\medskip
\sitocline{Supplementary Figure~\ref*{fig:size_matching}: Data-center size-matching sensitivity (scenario group G8)}{fig:size_matching}
\sitocline{Supplementary Figure~\ref*{fig:cooling_response}: Cooling-side response of the absorption configuration}{fig:cooling_response}
\medskip
\sitocline{Supplementary Table~\ref*{tab:size_matching}: Data-center size-matching results for Case~2}{tab:size_matching}
\sitocline{Supplementary Table~\ref*{tab:secondary_kpis_cost_carbon}: Secondary cost and carbon endpoints}{tab:secondary_kpis_cost_carbon}
\sitocline{Supplementary Table~\ref*{tab:secondary_kpis_water}: Water-footprint endpoints}{tab:secondary_kpis_water}
\sitocline{Supplementary Table~\ref*{tab:value_decomp}: Marginal annual value of adding the absorption chiller}{tab:value_decomp}
\sitocline{Supplementary Table~\ref*{tab:market_years}: Market inputs and on-site economics by operating year}{tab:market_years}
\sitocline{Supplementary Table~\ref*{tab:endpoint_factors}: Water-consumption and embodied-energy factors}{tab:endpoint_factors}
\sitocline{Supplementary Table~\ref*{tab:data}: Input data sources, baseline values and units}{tab:data}
\sitocline{Supplementary Table~\ref*{tab:equipment_tech}: Equipment technical parameters}{tab:equipment_tech}

\sitocline{Supplementary Table~\ref*{tab:robustness}: Operating and capital-recovery sensitivities}{tab:robustness}
}
\clearpage
\subsection*{Supplementary Note 1: Data-center size-matching results}
\begin{figure}[pos=tp]
\centering
\includegraphics[width=\linewidth]{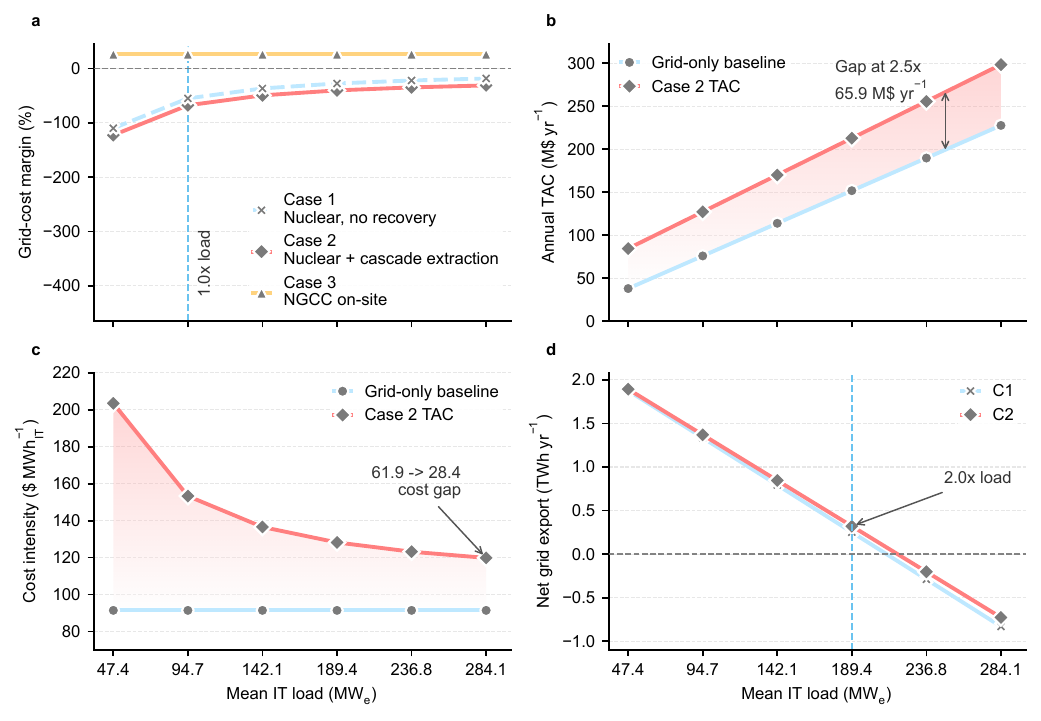}
\caption{Data-center size-matching sensitivity (scenario group G8). The IT-load trace and the data-center-side cooling and NGCC capacities are scaled from 0.5$\times$ to 3.0$\times$ while the BWRX-300 reactor remains fixed at 300~MW$_\mathrm{e}$ net; the PCC is held at 300~MW below 1$\times$ and scales upward above 1$\times$. (a)~Grid-cost margin for the on-site cases against the matching grid-only baseline. (b)~Case~2 TAC against the matching grid-only TAC; the shaded area is the annual dollar gap. (c)~IT-energy-normalized cost intensity; the shaded area is the excess Case~2 cost over grid supply. (d)~Net grid export for the two nuclear cases; positive values are merchant export, negative values net grid import. Arrow labels mark the 2.5$\times$ load point in panel~(b), the 1.0$\times$-to-3.0$\times$ cost-gap span in panel~(c) and the 2.0$\times$ load point in panel~(d); the zero-net-export crossing lies between 2.0$\times$ and 2.5$\times$.}
\label{fig:size_matching}
\end{figure}

\phantomsection\label{note:size}
Supplementary Figure~\ref{fig:size_matching} and Supplementary Table~\ref{tab:size_matching} report the size-matching sweep (scenario group G8). Holding the BWRX-300 reactor fixed at 300~MW$_\mathrm{e}$ net while the IT load and the data-center-side cooling and NGCC capacities scale from 0.5$\times$ to 3.0$\times$ moves Case~2 from a merchant-export surplus to a net import position, with the crossing between 2.0$\times$ and 2.5$\times$. The annual dollar gap to grid supply widens with scale, as the larger campus buys more grid power during scheduled outages; the IT-energy-normalized cost nevertheless falls from \$153.3 to \$119.9~MWh$_\mathrm{IT}^{-1}$ and the excess over grid supply from \$61.9 to \$28.4~MWh$_\mathrm{IT}^{-1}$ between 1.0$\times$ and 3.0$\times$. At baseline, 1.45~TWh of gross exports less 0.08~TWh of imports gives 1.37~TWh of net exports. At 2.0$\times$ the site still exports 0.32~TWh, with a matched margin of $-40.2\%$.

\begin{table}[pos=t]
\centering
\caption{Data-center size-matching results for Case~2 (scenario group G8) at the 2023 ATB-Mid baseline.}
\label{tab:size_matching}
\renewcommand{\arraystretch}{1.08}
\small
\begin{tabularx}{\linewidth}{@{}L{0.34\linewidth}*{6}{R}@{}}
\toprule
Load multiplier & 0.5 & 1.0 & 1.5 & 2.0 & 2.5 & 3.0 \\
\midrule
Case~2 TAC (\$M~yr$^{-1}$) & 84.4 & 127.2 & 170.0 & 212.8 & 255.5 & 298.3 \\
Grid-only TAC (\$M~yr$^{-1}$)                & 37.9 & 75.9 & 113.8 & 151.7 & 189.7 & 227.6 \\
Annual gap (\$M~yr$^{-1}$) & 46.5 & 51.3 & 56.2 & 61.0 & 65.9 & 70.7 \\
Grid-cost margin $M$ (\%) & \textminus122.6 & \textminus67.7 & \textminus49.4 & \textminus40.2 & \textminus34.7 & \textminus31.1 \\
Net grid export (TWh~yr$^{-1}$) & +1.89 & +1.37 & +0.85 & +0.32 & \textminus0.20 & \textminus0.73 \\
Cost intensity (\$~MWh$_\mathrm{IT}^{-1}$) & 203.5 & 153.3 & 136.6 & 128.2 & 123.2 & 119.9 \\
Excess over grid (\$~MWh$_\mathrm{IT}^{-1}$) & 112.1 & 61.9 & 45.1 & 36.8 & 31.8 & 28.4 \\
\bottomrule
\end{tabularx}
\begin{flushleft}
\footnotesize \textit{Note:} The reactor is held at 300~MW$_\mathrm{e}$ net while the IT load and the data-center-side cooling and NGCC capacities scale with the multiplier; the PCC remains 300~MW below 1$\times$ and scales upward above 1$\times$. $M$ is the grid-cost margin against the matched grid-only baseline; net export is annual grid sales minus purchases. Rows are rounded independently, so the annual gap may differ from the difference of the rounded totals by 0.1.
\end{flushleft}
\end{table}

\subsection*{Supplementary Note 2: Secondary cost, carbon and water endpoints}\phantomsection\label{note:secondary}
The run-level secondary endpoints at the 2023 ATB-Mid baseline complement the grid-cost margin reported in the main article. Supplementary Table~\ref{tab:secondary_kpis_cost_carbon} lists the cost and carbon endpoints, and Supplementary Table~\ref{tab:secondary_kpis_water} the water-footprint endpoints, for the four configurations.

\begin{table}[pos=t]
\centering
\caption{Secondary cost, carbon and energy-utilization endpoints at the 2023 ATB-Mid baseline.}
\label{tab:secondary_kpis_cost_carbon}
\renewcommand{\arraystretch}{1.08}
\small
\begin{tabularx}{\linewidth}{@{}L{0.26\linewidth}*{6}{R}@{}}
\toprule
Configuration & NLCS & LCOC & CO$_2$ balance & EPBT & $\eta_\mathrm{el}$ & $U$ \\
              & (\$~MWh$_\mathrm{e}^{-1}$) & (\$~MWh$_\mathrm{c}^{-1}$) & (kt~yr$^{-1}$) & (yr) & (\%) & (\%) \\
\midrule
Case~0, grid-only & 91.4 & 24.1 & +363 & -- & -- & -- \\
Case~1, SMR & 141.7 & -- & \textminus422 & 0.56 & 34.5 & 34.5 \\
Case~2, SMR + absorption & 153.3 & -- & \textminus435 & 0.57 & 33.8 & 37.5 \\
Case~3, NGCC & 67.0 & -- & +486 & 0.18 & -- & -- \\
\bottomrule
\end{tabularx}
\begin{flushleft}
\footnotesize \textit{Note:} NLCS is net TAC---including cooling capital, carbon cost, the production tax credit and export netting---divided by IT energy delivered. Net CO$_2$ credits merchant exports at the ERCOT hourly average emission factor. $\eta_\mathrm{el}$ is annual net electricity divided by reactor heat; $U$ adds useful extracted heat to the numerator at the reactor--steam-outlet boundary (Supplementary Note~10). Neither is an exergy efficiency. NGCC EPBT uses an illustrative embodied-energy assumption (Supplementary Note~7). Dashes (--) mark endpoints not evaluated for that configuration.
\end{flushleft}
\end{table}

\begin{table}[pos=t]
\centering
\caption{Water-footprint endpoints at the 2023 ATB-Mid baseline.}
\label{tab:secondary_kpis_water}
\renewcommand{\arraystretch}{1.08}
\small
\begin{tabularx}{\linewidth}{@{}L{0.34\linewidth}*{3}{R}@{}}
\toprule
Configuration & Direct site & Indirect generation & Total water \\
              & (L~MWh$_\mathrm{e}^{-1}$) & (L~MWh$_\mathrm{e}^{-1}$) & (L~MWh$_\mathrm{e}^{-1}$) \\
\midrule
Case~0, grid-only & 111 & 1{,}853 & 1{,}964 \\
Case~1, SMR & 111 & 7{,}547 & 7{,}658 \\
Case~2, SMR + absorption & 141 & 7{,}396 & 7{,}537 \\
Case~3, NGCC & 111 & 1{,}018 & 1{,}129 \\
\bottomrule
\end{tabularx}
\begin{flushleft}
\footnotesize \textit{Note:} Direct site water covers data-center cooling-system consumption; generation water covers on-site electricity production and imported electricity.
\end{flushleft}
\end{table}

\subsection*{Supplementary Note 3: Marginal value of absorption cooling}
\begin{figure}[pos=tp]
\centering
\includegraphics[width=\linewidth]{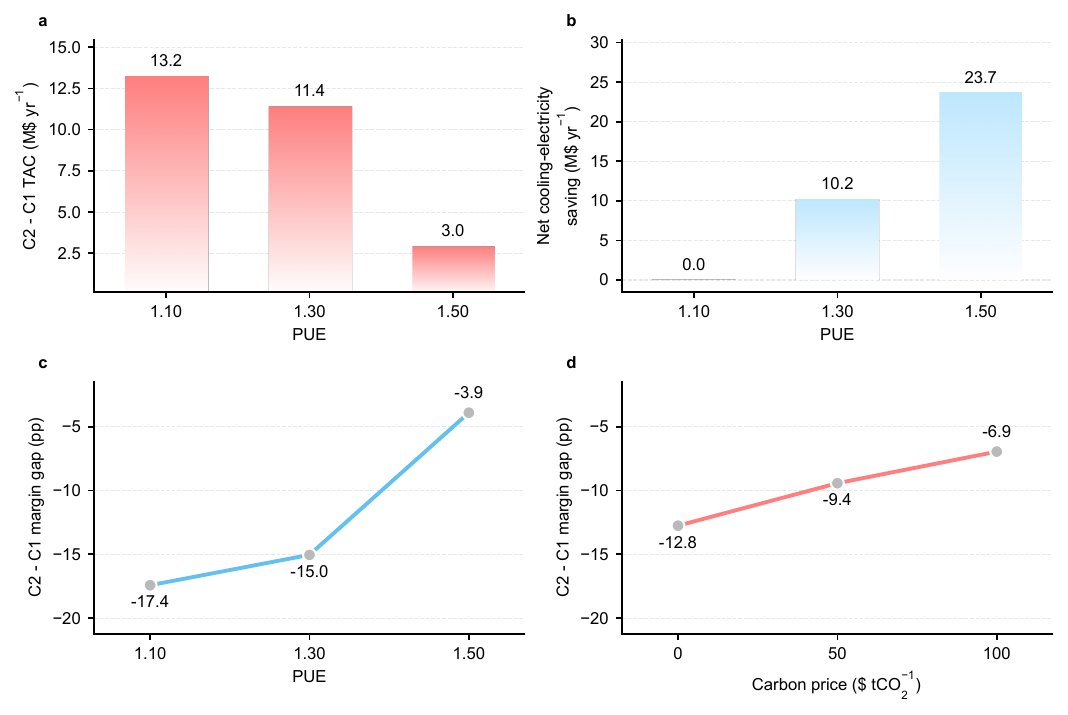}
\caption{Cooling-side response of the absorption configuration at the 2023 ATB-Mid baseline. (a)~Case~2 minus Case~1 TAC difference across full-load PUE; positive is a cogeneration penalty, negative a saving. (b)~Net cooling-electricity saving from absorption, measured as gross avoided VCC electricity cost less the VCC dispatch retained in Case~2, before fixed absorption capital and turbine-opportunity costs. (c)~The same difference expressed as a grid-cost-margin gap across full-load PUE. (d)~The margin gap across carbon price. At the baseline absorption capital of \$750~kW$_\mathrm{c}^{-1}$ and with the generation-indexed Section~45Y credit, Case~2 remains more expensive than Case~1 throughout the PUE sweep.}
\label{fig:cooling_response}
\end{figure}

\phantomsection\label{note:absvalue}
Supplementary Figure~\ref{fig:cooling_response} and Supplementary Table~\ref{tab:value_decomp} report the marginal value of adding the absorption chiller to the reactor-only plant (Case~2 minus Case~1) at the 2023 ATB-Mid baseline and across the cooling-efficiency sweep. Gross avoided vapor-compression electricity grows with full-load PUE, but at the baseline absorption capital of \$750~kW$_\mathrm{c}^{-1}$ the penalty never turns into a saving inside the swept range. Against a highly efficient electric chiller (PUE~1.10) the absorber never dispatches, and the gap is its idle \$13.2~M~yr$^{-1}$ capital and fixed-cost burden; at PUE~1.50 the gap narrows to \$3.0~M~yr$^{-1}$ but remains positive.

The penalty depends only weakly on the market year: about \$9.8 and \$9.7~M~yr$^{-1}$ in 2022 and 2023, and \$12.5~M~yr$^{-1}$ in the cheap-grid 2024. Higher prices raise the avoided electric cooling and the value of the forfeited generation together, so the two effects largely cancel. The extraction penalty matters far more. Sweeping $\alpha_\mathrm{w}$ over 0.08--0.25 at the baseline moves the gap from \$3.2 to \$11.5~M~yr$^{-1}$ and the absorption share of cooling from 91.8\% to 11.7\%: dispatch is highly sensitive to the steam opportunity cost, yet the net value of absorption stays negative across the tested coefficient range at current absorption capital costs.

Absorption capital is a dispatch-neutral annuity. It enters the objective of Eq.~\eqref{eq:tac} only as $\mathrm{CRF}_\mathrm{a}\,\mathrm{CAPEX}_\mathrm{a}S_\mathrm{a}$ with the absorber fixed at $S_\mathrm{a}=160$~MW$_\mathrm{c}$, appears in no constraint, and has no counterpart in Case~1, which installs no absorber. The optimal dispatch of both cases is therefore independent of it, and the Case~2 minus Case~1 gap is exactly linear in $\mathrm{CAPEX}_\mathrm{a}$ with slope $\mathrm{CRF}_\mathrm{a}S_\mathrm{a}=0.0835\times160{,}000 = \$13.4$~M~yr$^{-1}$ per \$1{,}000~kW$_\mathrm{c}^{-1}$. Two checks confirm the coefficient: at the \$750~kW$_\mathrm{c}^{-1}$ baseline it gives \$10.0~M~yr$^{-1}$ of capital recovery, which, with the \$20~kW$_\mathrm{c}^{-1}$~yr$^{-1}$ fixed O\&M (\$3.2~M~yr$^{-1}$), reproduces the \$13.2~M~yr$^{-1}$ row of Supplementary Table~\ref{tab:value_decomp}; and across the surveyed \$450--1{,}200~kW$_\mathrm{c}^{-1}$ range it moves Case~2 total annualized cost by \$10.0~M~yr$^{-1}$, or 13.2\% of the \$75.9~M~yr$^{-1}$ grid-only baseline, which is the 13.2-percentage-point absorption span of Figure~\ref{fig:boundary_atlas}d.

The break-even absorption capital reported in the last row of Supplementary Table~\ref{tab:value_decomp} follows as $\$750 - \Delta\mathrm{TAC}/(\mathrm{CRF}_\mathrm{a}S_\mathrm{a})$ and requires no additional runs. Only two of the four columns admit a non-negative solution. At the baseline the requirement is \$26~kW$_\mathrm{c}^{-1}$, below the \$450~kW$_\mathrm{c}^{-1}$ Bare-Low anchor by a factor of seventeen and below the \$250~kW$_\mathrm{c}^{-1}$ vapor-compression chiller capital benchmark; a free absorber would undercut the reactor-only plant by only \$0.34~M~yr$^{-1}$. At full-load PUE~1.50 the requirement is \$529~kW$_\mathrm{c}^{-1}$, inside the surveyed range, and a Bare-Low absorber would turn the \$3.0~M~yr$^{-1}$ penalty into a \$1.1~M~yr$^{-1}$ saving. At PUE~1.10 and 1.30, the operating benefit after lost generation, forgone PTC and residual operating costs is below the absorber's fixed O\&M. The cost gap therefore remains positive even at zero absorption capital cost. These capital thresholds apply to the fixed installed absorption capacity; absorber sizing is not optimized.

\begin{table}[pos=t]
\centering
\caption{Marginal annual value of adding the absorption chiller to a reactor-only plant (Case~2 minus Case~1), at the 2023 ATB-Mid baseline (full-load PUE~1.35) and across the cooling-efficiency sweep.}
\label{tab:value_decomp}
\renewcommand{\arraystretch}{1.08}
\small
\begin{tabularx}{\linewidth}{@{}L{0.34\linewidth}*{4}{R}@{}}
\toprule
Component & Baseline & PUE~1.10 & PUE~1.30 & PUE~1.50 \\
\midrule
Gross avoided VCC electricity & +17.2 & +4.9 & +14.8 & +24.7 \\
VCC dispatch retained in Case~2 & \textminus3.6 & \textminus4.9 & \textminus4.6 & \textminus1.1 \\
Absorption capital and fixed O\&M & \textminus13.2 & \textminus13.2 & \textminus13.2 & \textminus13.2 \\
Willans extraction opportunity cost & \textminus7.7 & 0.0 & \textminus6.7 & \textminus9.5 \\
Production tax credit forgone & \textminus1.0 & 0.0 & \textminus0.5 & \textminus2.4 \\
Residual reconciliation & \textminus1.4 & 0.0 & \textminus1.3 & \textminus1.5 \\
\midrule
Net value of absorption & \textminus9.7 & \textminus13.2 & \textminus11.4 & \textminus3.0 \\
Direct site-water externality (memo) & \textminus0.01 & 0.00 & \textminus0.01 & \textminus0.03 \\
Break-even absorption capital (memo) & 26 & n.a. & n.a. & 529 \\
\bottomrule
\end{tabularx}
\begin{flushleft}
\footnotesize \textit{Note:} All entries are \$M~yr$^{-1}$ except the break-even row, which is \$~kW$_\mathrm{c}^{-1}$. Annual cost components are shown to one decimal; the small water-cost memo retains two decimals. Positive values reduce Case~2 cost relative to Case~1; negative values increase it. Components are rounded independently and may not sum exactly to the net value. The direct site-water increment carries no price in the TAC objective, so it is reported as a memo externality outside the component sum and the residual. The break-even absorption capital is the installed cost at which the net value reaches zero, obtained in closed form because absorption capital is dispatch-neutral. In columns marked ``n.a.'', the operating benefit after lost generation, forgone PTC and residual operating costs is below fixed O\&M, so no non-negative absorption capital cost closes the gap.
\end{flushleft}
\end{table}

\subsection*{Supplementary Note 4: Cogeneration dispatch and storage detail}\phantomsection\label{note:dispatch}
This note records the dispatch and storage detail behind Sections~\ref{sec:results:cooling} and~\ref{sec:results:frontier} of the main article. During the 701-hour aggregate outage the reactor and the absorber are both offline---the absorber's annual maintenance is scheduled inside the window---and the site imports an average of 121.0~MW of grid power while the electric chiller carries 72.9~GWh$_\mathrm{c}$ of cooling. In the islanded gas case (Case~3) the 2023 day-ahead price falls below the NGCC full-load running-cost benchmark of \$20.8~MWh$_\mathrm{e}^{-1}$, evaluated at the annual-mean fuel price, in 41.9\% of hours; unable to import, the plant cannot arbitrage them, and the comparison against grid supply is made on annual cost. The lithium-ion battery sweep (scenario group G3) raises total annualized cost by \$0.76--0.77~M~yr$^{-1}$ across the nuclear cases at the two-hour utility-scale capital cost of \$529~kWh$^{-1}$: arbitrage value no longer covers the battery's annualized capital cost. This corresponds to a reduction of 1.0 percentage points in the grid-cost margin, the smallest span of the eight sensitivity axes, and it leaves the configuration ranking unchanged; battery deployment is treated as a secondary sensitivity throughout.

\subsection*{Supplementary Note 5: Market inputs by operating year}\phantomsection\label{note:market}
The three modeled years bracket the recent range of ERCOT and natural-gas market conditions and support the choice of 2023 as the baseline operating year (Supplementary Table~\ref{tab:market_years}). The year labels describe annual-mean day-ahead LMP and delivered gas price; short-duration price spikes can occur within each regime.

Because reactor capital enters the objective only as a dispatch-neutral annualized term, the annual cost separates exactly as $\mathrm{TAC}=K(\mathrm{capital})+D(\mathrm{year})$, and evaluating the nth-of-a-kind and \$5{,}000~kW$_\mathrm{e}^{-1}$ anchors in the 2022 and 2024 regimes (scenario group G9) uses this exact fixed-capital adjustment to the optimized annual dispatch. In the high-price 2022 regime the nth-of-a-kind margins rise to $+121.5\%$ for Case~1 and $+109.9\%$ for Case~2, and the \$5{,}000~kW$_\mathrm{e}^{-1}$ anchor already returns $+51.3\%$ and $+39.8\%$. In the low-price 2024 regime the same nth-of-a-kind anchor falls to $-0.5\%$ for Case~1 and $-34.8\%$ for Case~2, and the \$5{,}000~kW$_\mathrm{e}^{-1}$ anchor to $-164.4\%$ and $-198.7\%$. Each margin is quoted against that year's own grid-only baseline, which itself falls from \$85.1 to \$36.5~M~yr$^{-1}$ between 2022 and 2024, so part of the swing is a denominator effect; the \$5{,}000~kW$_\mathrm{e}^{-1}$ parity threshold reported in the main article is specific to the 2023 price year.

\begin{table}[pos=t]
\centering
\caption{Market inputs and on-site economics by operating year.}
\label{tab:market_years}
\renewcommand{\arraystretch}{1.08}
\small
\begin{tabularx}{\linewidth}{@{}L{0.34\linewidth}C{0.24\linewidth}*{3}{R}@{}}
\toprule
Quantity & Unit & 2022 & 2023 & 2024 \\
\midrule
ERCOT day-ahead LMP       & \$~MWh$_\mathrm{e}^{-1}$ & 69.9 & 57.3 & 28.3 \\
Delivered natural gas     & \$~MMBtu$^{-1}$          & 6.53 & 2.63  & 2.35 \\
Grid-only (Case~0) TAC    & \$M~yr$^{-1}$            & 85.1 & 75.9  & 36.5 \\
NGCC (Case~3) TAC         & \$M~yr$^{-1}$            & 87.0 & 55.6  & 53.5 \\
NGCC grid-cost margin $M$ & \%                       & \textminus2.2 & +26.7 & \textminus46.8 \\
\bottomrule
\end{tabularx}
\begin{flushleft}
\footnotesize \textit{Note:} ERCOT LMP is the annual-mean Houston Hub day-ahead price; delivered gas is the annual-mean Henry Hub price plus the Houston Ship Channel basis. $M$ is positive when the configuration is cheaper than grid procurement. Margins are computed from unrounded totals and may differ slightly from the ratio of the rounded values shown.
\end{flushleft}
\end{table}

\subsection*{Supplementary Note 6: Exogenous data sources and provenance}\phantomsection\label{note:provenance}
The time-series inputs are assembled from public sources and aligned to an 8760-hour annual horizon. ERCOT day-ahead locational marginal prices for the Houston Hub node (HB\_HOUSTON) are retrieved through the gridstatus.io public data interface \citep{gridstatus}. Hourly average grid carbon intensities are derived from the EIA-930 Hourly Electric Grid Monitor generation mix \citep{eia930}, weighted by UNECE 2022 lifecycle emission factors for each technology \citep{unece2022lca}. Houston wet-bulb temperature is taken from the Open-Meteo historical-weather reanalysis archive \citep{openmeteo2023}. The IT-load trace is derived from the Vercellino et al.\ 2026 NLR Data Catalog colocation profile \citep{vercellino2026aiworkload}. The 10~MW source profile is not a field measurement: it is the output of a year-long, one-minute discrete-event simulation whose facility-utilization distributions are adapted from the Shanghai Artificial Intelligence Laboratory ACME job traces and whose seasonality follows monthly distributions from the NLR Kestrel high-performance computing system. We take the mean-utilization scenario whose scaled annual mean matches the 94.7~MW$_\mathrm{e}$ reported here, resample it to hourly resolution, scale it by 20$\times$ to represent a 200~MW facility, and circularly shift it by whole days so that day-of-week aligns with each market year. Delivered natural gas is the EIA Henry Hub daily spot price \citep{eiaHenryHub} plus a fixed Houston Ship Channel basis.

Cost anchors follow the NLR 2024 Annual Technology Baseline \citep{atb2024nuclear} for the ATB-Mid reactor capital case, the NGCC cost block, the reactor and NGCC fixed O\&M, and the utility-scale battery storage case. The FOAK and NOAK reactor anchors are the Ontario Power Generation Darlington unit budget on an SMR-unit-only basis and the GE-Hitachi vendor target, respectively. The reactor capacity factor is the EIA fleet average \citep{eia2025capacityfactors}, the nuclear fuel cost follows the companion nuclear--hydrogen study \citep{li2026nuclearH2}, absorption-chiller capital costs are bracketed from U.S.\ Department of Energy combined-heat-and-power fact-sheet data \citep{doe2017chpAbsorption} and publicly available vendor literature, and the vapor-compression chiller capital cost is a typical water-cooled centrifugal value from ASHRAE engineering practice.

The Section~45Y clean-electricity production tax credit follows the enacting statute \citep{usc26_45y,ira2022}, the calendar-2025 inflation adjustment \citep{fr2025_45y} and U.S.\ Internal Revenue Service guidance \citep{irs45y}. No proprietary or restricted-access data are used, and the audit trail linking each raw file to its processed model input is included in the project repository.

\subsection*{Supplementary Note 7: Secondary endpoint definitions}\phantomsection\label{note:endpoints}

This note defines the secondary endpoints reported in Supplementary Tables~\ref{tab:secondary_kpis_cost_carbon} and~\ref{tab:secondary_kpis_water} and quoted in the main text, together with their system boundaries and data sources. The endpoints use the same solved dispatch as the total annualized cost (TAC) of Eq.~\eqref{eq:tac}. Water consumption and energy payback additionally use the factors listed in Supplementary Table~\ref{tab:endpoint_factors}. Throughout, $E_\mathrm{IT}=\sum_t P_\mathrm{IT}(t)\,\Delta t = 829.7$~GWh$_\mathrm{e}$~yr$^{-1}$ is the annual IT energy delivered, identical in every case and, apart from the G8 size sweep, every scenario including the energy-matched constant-load sensitivity, and $E_\mathrm{gen}$ denotes the annual net on-site generation (turbine net output $\sum_t P_\mathrm{tn}(t)\Delta t$ for the nuclear cases; NGCC output $\sum_t P_\mathrm{ng}(t)\Delta t$ for Case~3).

\runinhead{Net levelized cost of IT supply (NLCS)}
The NLCS divides the full TAC by the IT energy delivered,
\begin{equation}
\label{eq:nlcs}
\mathrm{NLCS}_c \;=\; \frac{\mathrm{TAC}_c}{E_\mathrm{IT}},
\end{equation}
so its numerator inherits every term of Eq.~\eqref{eq:tac}: annualized capital of all installed blocks including the cooling plant, fixed and variable operations and maintenance, fuel, the net grid exchange $\Phi_\mathrm{grid}$ (merchant-export revenue enters with a negative sign), the carbon cost when priced, and the Section~45Y credit $-\Phi_\mathrm{ptc}$. The NLCS is therefore a cost of serving the IT load and should not be read as a levelized cost of electricity generation: for the export-heavy nuclear cases it nets roughly 1.42--1.45~TWh~yr$^{-1}$ of merchant sales against the cost of a 300~MW$_\mathrm{e}$ plant serving a 94.7~MW$_\mathrm{e}$ average load. Because $E_\mathrm{IT}$ is identical across cases, the NLCS ranking is by construction the TAC ranking rescaled ($\mathrm{NLCS}_c = \mathrm{TAC}_c/E_\mathrm{IT}$, so $M_c = 1-\mathrm{NLCS}_c/\mathrm{NLCS}_0$), and it is reported for interpretability rather than as an independent result.

\runinhead{Levelized cost of cooling (LCOC)}
For the grid-only Case~0 the cooling-attributable cost separates cleanly, and
\begin{equation}
\label{eq:lcoc}
\mathrm{LCOC}_0 \;=\; \frac{\mathrm{CRF}_\mathrm{v}\,\mathrm{CAPEX}_\mathrm{v} S_\mathrm{v} + \mathrm{FOM}_\mathrm{v} S_\mathrm{v} + \sum_t \big[v_\mathrm{v}\, Q_\mathrm{v}(t) + \pi_\mathrm{g}(t)\, P_\mathrm{v}(t)\big]\Delta t}{\sum_t Q_\mathrm{v}(t)\,\Delta t},
\end{equation}
i.e., the vapor-compression chiller's annualized capital, fixed and variable O\&M, plus the grid cost of the chiller's own electricity, divided by the chilled water delivered ($\sum_t Q_\mathrm{v}(t)\Delta t = 921.9$~GWh$_\mathrm{c}$~yr$^{-1}$). In Cases~1--3 the generation assets serve the IT and cooling loads jointly, so assigning shared capital and fuel to the cooling stream would require an arbitrary allocation rule; rather than impose one, the LCOC is reported only where it is allocation-free (Case~0), and cross-case cooling economics are carried by the TAC and the grid-cost margin.

\runinhead{Energy payback time (EPBT)}
Following the life-cycle energy convention for power plants \citep{lenzen2008nuclear}, the EPBT divides the cumulative primary-energy demand of plant construction and decommissioning, expressed as electricity-equivalent, by the plant's annual net generation:
\begin{equation}
\label{eq:epbt}
\mathrm{EPBT} \;=\; \frac{e_k\, S_k}{E_\mathrm{gen}},
\end{equation}
with $e_k$ the embodied-energy intensity per unit of installed capacity and $S_k$ the installed rating. The nuclear factor $e_\mathrm{rx}=4.5$~MWh$_\mathrm{e}$-eq~kW$_\mathrm{e}^{-1}$ is the light-water-reactor midpoint of the Lenzen review \citep{lenzen2008nuclear}, applied to the 300~MW$_\mathrm{e}$ net rating; the NGCC factor $e_\mathrm{ng}=1.0$~MWh$_\mathrm{e}$-eq~kW$_\mathrm{e}^{-1}$ is an illustrative screening assumption applied to the 200~MW$_\mathrm{e}$ block. The NGCC EPBT is conditional on this assumed factor. The denominator is the plant's full net generation including merchant exports, so the EPBT characterizes the generating asset itself rather than the IT share of its output; it is undefined for the grid-only Case~0, which amortizes no on-site plant. Nuclear-fuel-cycle energy is not included in the numerator; its operating-cost counterpart enters the TAC through the fuel price of Eq.~\eqref{eq:fuel_cost} \citep{li2026nuclearH2}.

\runinhead{Water consumption}
The water endpoint separates on-site cooling-tower makeup from upstream generation water and reports both per unit of IT energy delivered:
\begin{align}
W_\mathrm{dir} \;=\;& \frac{w_\mathrm{v} \sum_t Q_\mathrm{v}(t)\Delta t \;+\; w_\mathrm{a} \sum_t Q_\mathrm{a}(t)\Delta t}{E_\mathrm{IT}}, \label{eq:water_direct}\\
W_\mathrm{ind} \;=\;& \frac{w_\mathrm{rx} \sum_t P_\mathrm{tn}(t)\Delta t + w_\mathrm{ng} \sum_t P_\mathrm{ng}(t)\Delta t + w_\mathrm{g} \sum_t P_\mathrm{g}^{+}(t)\Delta t}{E_\mathrm{IT}}, \label{eq:water_indirect}\\
W_\mathrm{tot} \;=\;& W_\mathrm{dir} + W_\mathrm{ind}. \label{eq:water_total}
\end{align}
with the consumption factors $w$ listed in Supplementary Table~\ref{tab:endpoint_factors}. All factors are water \emph{consumption} (withdrawal net of return flow). The generation factors are per MWh$_\mathrm{e}$ \emph{generated} while the denominator is per MWh$_\mathrm{e}$ of IT energy \emph{delivered}: the nuclear term charges the full turbine output including the roughly 1.42--1.45~TWh~yr$^{-1}$ of merchant exports, because the cooling tower consumes that water at the site regardless of where the electricity is used. Unlike the carbon accounting of Eq.~\eqref{eq:carbon_cost}, this water inventory does not deduct an export credit. It attributes water consumption from all on-site generation to the campus, including generation sold to the grid. On this basis, Case~1 reports 7{,}658~L~MWh$_\mathrm{e}^{-1}$, compared with 1{,}964~L~MWh$_\mathrm{e}^{-1}$ for grid supply. The absorption factor resolves the trade documented in the main text: per MWh$_\mathrm{c}$ delivered, an absorption chiller rejects its cooling duty plus its driving heat, $(1+1/\mathrm{COP}_\mathrm{a})\approx 1.83$ times the delivered cooling at the Houston annual-mean $\mathrm{COP}_\mathrm{a}\approx1.2$, against roughly one for the electric chiller once its compressor work is neglected, so $w_\mathrm{a} = w_\mathrm{v}\,(1+1/1.2) = 0.183$~L~kWh$_\mathrm{c}^{-1}$. The direct-site tier is also the basis of the memo externality in Supplementary Table~\ref{tab:value_decomp}, which prices the Case~2 minus Case~1 increment of $W_\mathrm{dir} E_\mathrm{IT}$ at \$0.50~m$^{-3}$, an illustrative water-price assumption; at \$0.01~M~yr$^{-1}$ it is three orders of magnitude below the TAC terms.

\begin{table}[pos=t]
\centering
\caption{Water-consumption factors and embodied-energy intensities used in the secondary endpoints. VCC, vapor-compression chiller; NGCC, natural gas combined cycle. Generation factors are medians for recirculating (cooling-tower) plants; consumption basis throughout.}
\label{tab:endpoint_factors}
\renewcommand{\arraystretch}{1.08}
\begin{tabularx}{\linewidth}{@{}L{0.11\linewidth}L{0.34\linewidth}L{0.21\linewidth}Y@{}}
\toprule
Symbol & Quantity & Value & Source \\
\midrule
$w_\mathrm{v}$  & VCC site water per cooling delivered & 0.10~L~kWh$_\mathrm{c}^{-1}$ & authors' estimate for hybrid cooling-tower duty; \citet{macknick2012water} covers generation only \\
$w_\mathrm{a}$  & absorption site water per cooling delivered & 0.183~L~kWh$_\mathrm{c}^{-1}$ & $w_\mathrm{v}(1+1/\mathrm{COP}_\mathrm{a})$ at annual-mean $\mathrm{COP}_\mathrm{a}=1.2$ \\
$w_\mathrm{rx}$ & nuclear generation, recirculating tower & 2.54~L~kWh$_\mathrm{e}^{-1}$ & \citet{macknick2012water}, median \\
$w_\mathrm{ng}$ & NGCC generation, recirculating tower & 0.78~L~kWh$_\mathrm{e}^{-1}$ & \citet{macknick2012water}, median \\
$w_\mathrm{g}$  & ERCOT grid blend & 1.42~L~kWh$_\mathrm{e}^{-1}$ & \citet{macknick2012water} factors weighted by the ERCOT mix \\
$e_\mathrm{rx}$ & nuclear embodied energy & 4.5~MWh$_\mathrm{e}$-eq~kW$_\mathrm{e}^{-1}$ & \citet{lenzen2008nuclear}, light-water midpoint \\
$e_\mathrm{ng}$ & NGCC embodied energy & 1.0~MWh$_\mathrm{e}$-eq~kW$_\mathrm{e}^{-1}$ & illustrative screening assumption \\
\bottomrule
\end{tabularx}
\end{table}

\subsection*{Supplementary Note 8: Production-tax-credit levelization robustness}\phantomsection\label{note:ptcwindow}
The statutory Section~45Y credit lasts ten years from commissioning. Levelizing it over the twenty-year reference window with the annuity ratio of Eq.~\eqref{eq:ptc} gives \$19.7~MWh$_\mathrm{e}^{-1}$ at the baseline WACC. Levelizing the same ten-year credit over the reactor's own forty-year amortization window instead gives \$15.5~MWh$_\mathrm{e}^{-1}$.

Re-evaluating the nuclear scenarios with dispatch optimized for that credit window reduces the baseline grid-cost margins by 13.5 and 13.2 percentage points for Cases~1 and~2, moves the Case~2 parity threshold from roughly \$5{,}250 to about \$4{,}790~kW$_\mathrm{e}^{-1}$, and moves the carbon-price crossovers from \$53.1 and \$64.0 to \$66.1 and \$76.1~tCO$_2^{-1}$, which remain inside the solved \$0--100~tCO$_2^{-1}$ range. The thresholds therefore depend on the levelization window. At the baseline, both nuclear configurations remain more expensive than grid supply, and Case~2 remains more expensive than Case~1. Construction-start timing, prevailing-wage compliance and future inflation adjustments would change the credit's effective value further.

Under the current Section~45Y construction-start phaseout for eligible non-wind/solar facilities, the applicable percentage is 100\% through 2033, 75\% in 2034, 50\% in 2035 and zero thereafter \citep{usc26_45y}. These timing rules are separate from the fixed calendar-2025 credit scenario evaluated here.

\subsection*{Supplementary Note 9: Model formulation detail and full input inventory}\phantomsection\label{note:formulation}
The following equations specify the constraints and cost terms summarized in Section~\ref{sec:methods:opt}. The input inventory and equipment technical parameters are listed at the end of this note.

\runinhead{Reactor block (Cases~1 and 2)}
The reactor is bounded, held offline through the scheduled outage window and ramp-limited by
\begin{align}
\underline{P}_\mathrm{rx} \;\le\; & P_\mathrm{rx}(t) \;\le\; \overline{P}_\mathrm{rx}, &\forall\, t \in \mathcal{T}\setminus\mathcal{T}_\mathrm{out}, \label{eq:rx_bounds}\\
P_\mathrm{rx}(t) \;=\; & 0, &\forall\, t \in \mathcal{T}_\mathrm{out}, \label{eq:rx_outage}\\
\big|P_\mathrm{rx}(t) - P_\mathrm{rx}(t-1)\big| \;\le\; & \Delta P_\mathrm{rx}^{\max}, &\forall\, t \ge 1 \text{ outside } \mathcal{T}_\mathrm{out}. \label{eq:rx_ramp}
\end{align}
The outage window comprises $\mathrm{round}\!\big((1-\overline{\mathrm{CF}})\cdot 8760\big) = 701$ contiguous hours beginning at hour 1752 of the trace (15~March; 14~March in the leap-year 2024 record), so a plant at full power in every online hour realizes $\overline{\mathrm{CF}}=0.92$ to within 0.003\%. Shutdown and startup at the window boundaries are managed procedures and are excluded from the normal load-following ramp limit.

\runinhead{Battery state of charge}
The battery follows
\begin{equation}
\label{eq:bess_soc}
E_\mathrm{B}(t) \;=\; (1-\rho_\mathrm{B})\,E_\mathrm{B}(t-1) + \big(\sqrt{\eta_\mathrm{rt}}\,B^{+}(t) - B^{-}(t)/\sqrt{\eta_\mathrm{rt}}\big)\Delta t,
\end{equation}
with the initial state of charge fixed at $E_\mathrm{B}(0)=0.5\,\mathrm{Cap}_\mathrm{E}$ and cyclically restored through $E_\mathrm{B}(|\mathcal{T}|)=E_\mathrm{B}(0)$.

\runinhead{Variable operating and fuel costs}
The variable operations-and-maintenance and nuclear fuel terms of Eq.~\eqref{eq:tac} are
\begin{align}
\mathrm{VOM} ={}&
  \sum_t \big[v_\mathrm{rx} P_\mathrm{tn}(t) + v_\mathrm{v} Q_\mathrm{v}(t) + v_\mathrm{a} Q_\mathrm{a}(t)\big]\Delta t
  \notag\\
& + \sum_t v_\mathrm{B}\big(B^{+}(t)+B^{-}(t)\big)\Delta t, \label{eq:vom}\\
\Phi_\mathrm{fuel} ={}& \pi_\mathrm{f} \sum_t P_\mathrm{rx}(t)\Delta t, \label{eq:fuel_cost}
\end{align}
with the coefficients $v_x$ listed in Table~\ref{tab:equipment} of the main article. For Case~3 the corresponding gas terms are
\begin{align}
\Phi_\mathrm{fuel}^\mathrm{ng} ={}& \sum_t \pi_\mathrm{ng}(t)\,h_\mathrm{ng}\,\phi_\mathrm{ng}\!\big(\ell(t)\big)\,P_\mathrm{ng}(t)\Delta t, \label{eq:ngcc_fuel_cost}\\
\Phi_\mathrm{CO_2}^\mathrm{ng} ={}&
  \pi_\mathrm{CO_2}\Theta \sum_t
  \gamma_\mathrm{ng}\,\phi_\mathrm{ng}\!\big(\ell(t)\big)\, P_\mathrm{ng}(t)\Delta t. \label{eq:ngcc_carbon_cost}
\end{align}

\runinhead{Capital recovery}
Each asset is annualized at $\mathrm{CRF}_k=i(1+i)^{N_k}/[(1+i)^{N_k}-1]$, evaluated at the capital-recovery periods of Table~\ref{tab:equipment} and the baseline WACC $i=6.7\%$; at 25~yr and 6.7\% this gives the $\mathrm{CRF}_\mathrm{a}=0.0835$ used in Supplementary Note~3.

\runinhead{Input inventory}
Supplementary Table~\ref{tab:data} lists every exogenous input with its baseline value, resolution and source; Supplementary Table~\ref{tab:equipment_tech} lists the equipment technical parameters carried into the constraints above. Leap-year raw files are trimmed to 8760 hourly records for cross-year comparability.

\begin{table}[pos=t]
\centering
\caption{Input data sources, baseline values and units. UNECE, United Nations Economic Commission for Europe.}
\label{tab:data}
\renewcommand{\arraystretch}{1.08}
\begin{tabularx}{\linewidth}{@{}L{0.20\linewidth}L{0.30\linewidth}C{0.10\linewidth}Y@{}}
\toprule
Quantity & Baseline value or range & Resolution & Source \\
\midrule
ERCOT day-ahead LMP, Houston Hub        & annual mean \$28.3--69.9~MWh$_\mathrm{e}^{-1}$ (2022--2024) & hourly & gridstatus.io \\
ERCOT average carbon intensity          & 320--350~g~CO$_2$~kWh$_\mathrm{e}^{-1}$ annual mean (hourly average emission factor) & hourly & EIA-930 + UNECE 2022 \\
Delivered natural gas (Henry Hub + basis) & \$2.35--6.53~MMBtu$^{-1}$ (annual mean) & daily, applied hourly & EIA Henry Hub spot \\
Houston wet-bulb temperature            & $-0.5$ to 28~$^\circ$C annual range (2023) & hourly & Open-Meteo historical \\
Data-center IT load                     & 142.4~MW$_\mathrm{e}$ peak; 94.7~MW$_\mathrm{e}$ mean & hourly & NLR Data Catalog colocation profile, scaled 20$\times$ \citep{vercellino2026aiworkload} \\
BWRX-300 capital cost                   & \$7{,}615~kW$_\mathrm{e}^{-1}$ ATB-Mid; \$14{,}700~kW$_\mathrm{e}^{-1}$ FOAK; \$2{,}250~kW$_\mathrm{e}^{-1}$ NOAK & scalar & NLR ATB 2024 \citep{atb2024nuclear} (ATB-Mid); OPG Darlington budget, SMR-unit basis (FOAK); GE-Hitachi vendor target (NOAK) \\
BWRX-300 fixed O\&M                     & \$121~kW$_\mathrm{e}^{-1}$~yr$^{-1}$ & scalar & NLR ATB 2024 \citep{atb2024nuclear} \\
BWRX-300 capacity factor                & 0.92 (fleet average) & scalar & EIA Electric Power Monthly Table 6.07.B \citep{eia2025capacityfactors} \\
Generic nuclear lifecycle emissions    & 12~g~CO$_2$-eq~kWh$_\mathrm{e}^{-1}$ & scalar & IPCC lifecycle median \citep{ipcc2014annexiii} \\
Absorption chiller capital cost         & \$450--1{,}200~kW$_\mathrm{c}^{-1}$ (\$750 baseline) & scalar & U.S. Department of Energy combined-heat-and-power fact sheet \citep{doe2017chpAbsorption} + vendor literature \\
VCC capital cost                        & \$250~kW$_\mathrm{c}^{-1}$ & scalar & American Society of Heating, Refrigerating and Air-Conditioning Engineers (ASHRAE) typical centrifugal value \\
Lithium-ion BESS capital cost           & \$529~kWh$^{-1}$ (2-hour all-in, overnight) & scalar & NLR ATB 2024 utility-scale battery storage \citep{atb2024nuclear} \\
NGCC capital cost                       & \$1{,}330~kW$_\mathrm{e}^{-1}$ & scalar & NLR ATB 2024 F-class ATB-Mid \citep{atb2024nuclear} \\
NGCC direct CO$_2$                      & 360~g~CO$_2$~kWh$_\mathrm{e}^{-1}$ + 60~g upstream CH$_4$ (CO$_2$-eq) & scalar & EIA + Alvarez et al.\ (2018) \\
Nuclear fuel cost                       & \$10~MWh$_\mathrm{e}^{-1}$ net-electric basis ($\approx$\$3.45~MWh$_\mathrm{th}^{-1}$) & scalar & Li et al.\ 2026 \citep{li2026nuclearH2} \\
WACC                                    & 6.7\% baseline; \{5, 6.7, 10\}\% in scenario group G7 & scalar & NLR ATB 2024 financial assumptions \citep{atb2024nuclear} \\
Asset capital-recovery period           & 40~yr reactor island (60/80~yr sensitivities); 25~yr chillers; 30~yr NGCC; 15~yr battery & scalar & study financial assumptions; reactor technical life is 60~yr \citep{atb2024nuclear,gehitachi2026bwrx300} \\
Clean-electricity PTC (Section 45Y)     & \$30~MWh$_\mathrm{e}^{-1}$ prevailing-wage (calendar year 2025), 10-yr window, levelized to \$19.7~MWh$_\mathrm{e}^{-1}$ & scalar & 26~U.S.C.~\S\,45Y; 90 Federal Register 2025-16249 \\
\bottomrule
\end{tabularx}
\end{table}

\begin{table}[pos=t]
\centering
\caption{Equipment technical parameters used in the optimization. Derived coefficients are rounded to three significant figures; calculations retain the full precision of the model inputs.}
\label{tab:equipment_tech}
\renewcommand{\arraystretch}{1.08}
\begin{tabularx}{\linewidth}{@{}L{0.26\linewidth}Y@{}}
\toprule
Asset & Technical parameters \\
\midrule
BWRX-300 reactor (Cases~1,~2) & 870~MW$_\mathrm{th}$, 300~MW$_\mathrm{e}$ net; min-load fraction 0.50; ramp 0.5\%~min$^{-1}$; capacity factor 0.92 represented by a 701-h aggregate outage from 15~March; fuel \$10~MWh$_\mathrm{e}^{-1}$ net-electric basis ($\approx$\$3.45~MWh$_\mathrm{th}^{-1}$); lifecycle emissions 12~g~CO$_2$-eq~kWh$_\mathrm{e}^{-1}$ \\
Cascaded steam turbine (Cases~1,~2) & Willans line $a_\mathrm{w}\approx0.391~\mathrm{MW}_\mathrm{e}~\mathrm{MW}_\mathrm{th}^{-1}$, $b_\mathrm{w}\approx24.2~\mathrm{MW}_\mathrm{e}$ (316~MW$_\mathrm{e}$ gross, $\eta_\mathrm{r}=316/870\approx0.363$); auxiliary fraction $16/316$; extraction penalty $\alpha_\mathrm{w}=0.20~\mathrm{MW}_\mathrm{e}~\mathrm{MW}_\mathrm{th}^{-1}$ from the 7-bar crossover heat balance \\
Double-effect absorption chiller (Case~2) & 160~MW$_\mathrm{c}$ nominal; steam-side generator capacity 145.5~MW$_\mathrm{th}$ (= nameplate / design COP) and delivered-cooling cap 160~MW$_\mathrm{c}$; nameplate cap COP 1.30; design $T_\mathrm{wb,0}=26\,^\circ$C, COP 1.10; derating 0.015~$^\circ\mathrm{C}^{-1}$; crystallization shutdown $T_\mathrm{cry}=34\,^\circ$C cooling-water inlet (31~$^\circ$C design + 3~K margin); tower approach 5~K; parasitic $\beta_\mathrm{a}=0.035~\mathrm{kW}_\mathrm{e}~\mathrm{kW}_\mathrm{c}^{-1}$ (pumps 0.020 + tower fans 0.015) \\
Vapor-compression chiller (Cases~0--3) & water-cooled centrifugal; design-point COP 3.2 at $T_\mathrm{wb}=26\,^\circ$C with hourly wet-bulb relief (slope 0.0136~K$^{-1}$, cap 1.18, mirroring the absorber's relative response); 160~MW$_\mathrm{c}$ in all cases \\
Lithium-ion BESS (G3) & 100~MWh, 50~MW (C/2, 2-hour); round-trip $\eta_\mathrm{rt}=0.85$; self-discharge 0.1\%~day$^{-1}$; state-of-charge bounds [0.05, 0.95] \\
NGCC (Case~3) & 200~MW$_\mathrm{e}$ F-class; HHV $\eta_\mathrm{hhv}=0.495$; direct CO$_2$ 360~g~kWh$_\mathrm{e}^{-1}$; upstream CH$_4$ 60~g~kWh$_\mathrm{e}^{-1}$ (CO$_2$-eq); Henry Hub + Houston Ship Channel basis \$0.10~MMBtu$^{-1}$ \\
ERCOT interconnection (Cases~0--2) & 300~MW PCC bidirectional; interconnection capital and delays excluded \\
\bottomrule
\end{tabularx}
\end{table}

\subsection*{Supplementary Note 10: Energy utilization and operating-assumption sensitivities}\phantomsection\label{note:robustness}

\runinhead{Energy-utilization boundary}
The annual net-electric efficiency and electric-plus-heat utilization in Supplementary Table~\ref{tab:secondary_kpis_cost_carbon} are
\begin{equation}
\eta_\mathrm{el}=\frac{\sum_t P_\mathrm{tn}(t)\Delta t}{\sum_t P_\mathrm{rx}(t)\Delta t},\qquad
U=\frac{\sum_t [P_\mathrm{tn}(t)+Q_\mathrm{ext}(t)]\Delta t}{\sum_t P_\mathrm{rx}(t)\Delta t}.
\label{eq:utilization}
\end{equation}
The boundary includes the reactor and power block up to the useful steam outlet to the absorber. Net electricity is measured after station auxiliaries; chiller electricity and delivered cooling are downstream and are not added again. Both ratios include exported electricity and describe plant energy utilization rather than campus-only consumption or exergy efficiency. Case~1 has $U=\eta_\mathrm{el}=34.5\%$. Case~2 has $\eta_\mathrm{el}=33.8\%$ and $U=37.5\%$. Both cases have 7.01~TWh$_\mathrm{th}$ of annual reactor output and \$24.2~million of fuel cost, so the modeled utilization improvement does not reduce fuel use.

\runinhead{Refueling interval}
The 701-hour baseline window represents aggregate annual unavailability rather than the manufacturer's 10--15-day refueling duration. To isolate refueling frequency, an illustrative 14-day (336-hour) refueling component occupies the first part of the window and 365 hours of other downtime occupy the remainder. A two-year comparison combines one baseline year with one year retaining only the same 365 hours of other reactor downtime; both use the same 2023 price, weather and IT records, and the absorber retains its 701-hour annual maintenance window. The resulting mean reactor availability is 93.9\%, compared with 92.0\% in the annual-refueling baseline. Mean TAC falls by \$1.7~million~yr$^{-1}$ in each nuclear case, to \$115.8 and \$125.5~million~yr$^{-1}$, while the absorption share remains 32.6\%. Neither reaches grid parity (Supplementary Table~\ref{tab:robustness}). This controlled separation of outage components tests the cycle-length effect; it is not a forecast of the manufacturer's operating availability or a stochastic outage model.

\runinhead{Capital-recovery period}
Changing the reactor capital-recovery period from 40 to 60 or 80 years changes only its fixed annuity. At 6.7\% WACC, these changes reduce annual cost by \$9.2 or \$11.5~million in each nuclear case. The PTC levelization window remains 20 years, and hourly dispatch is unchanged. Case~2 retains a \$9.7~million annual premium over Case~1 and remains above grid cost. This sensitivity is distinct from the re-optimized 40-year PTC window in Supplementary Note~8; potential life-extension refurbishment and changing maintenance costs are not estimated.

\runinhead{Constant IT load}
The four configurations are also evaluated with IT demand fixed at its annual mean of 94.7~MW, preserving the baseline annual IT energy of 829.7~GWh, weather, market prices and installed equipment. This removes hourly and seasonal load variability without increasing annual utilization. Grid-only TAC falls from \$75.9 to \$67.0~million~yr$^{-1}$, reflecting the changed coincidence of demand with prices and chiller performance. Cases~1 and~2 cost \$108.7 and \$119.6~million~yr$^{-1}$, and Case~3 costs \$55.2~million~yr$^{-1}$. Absorption supplies 24.4\% of cooling and adds \$10.9~million~yr$^{-1}$ relative to Case~1. The cost ordering is unchanged; the matched nuclear margins become more negative because the grid benchmark falls more strongly.

\begin{table}[pos=t]
\centering
\caption{Operating and capital-recovery sensitivities at the 2023 ATB-Mid baseline. Costs are annual or two-year-mean TAC; margins use the matching Case~0 load profile.}
\label{tab:robustness}
\small
\begin{tabularx}{\linewidth}{@{}Yrrrr@{}}
\toprule
Assumption & Case~1 TAC & Case~2 TAC & Case~1 $M$ & Case~2 $M$ \\
 & (\$M~yr$^{-1}$) & (\$M~yr$^{-1}$) & (\%) & (\%) \\
\midrule
Baseline & 117.5 & 127.2 & $-54.9$ & $-67.7$ \\
Two-year refueling, unchanged other downtime & 115.8 & 125.5 & $-52.6$ & $-65.4$ \\
60-year reactor capital recovery & 108.4 & 118.0 & $-42.8$ & $-55.6$ \\
80-year reactor capital recovery & 106.0 & 115.7 & $-39.8$ & $-52.5$ \\
Equal-energy constant IT load & 108.7 & 119.6 & $-62.3$ & $-78.6$ \\
\bottomrule
\end{tabularx}
\end{table}

\runinhead{Nuclear carbon-billing exemption}
The nuclear-emissions pricing sensitivity sets $\gamma_\mathrm{rx}^\mathrm{bill}=0$ in Eq.~\eqref{eq:carbon_cost}, with the reported lifecycle factor $\gamma_\mathrm{rx}=12$ unchanged. All other carbon terms, including export credits at the hourly average grid factor, are retained. At \$50~tCO$_2^{-1}$, re-optimization lowers annual cost by \$1.5~million for Case~1 and \$1.4~million for Case~2; at \$100~tCO$_2^{-1}$ the reductions are \$2.9 and \$2.8~million. Linear interpolation of the solved TAC differences gives grid-parity prices of approximately \$51.2 and \$61.9~tCO$_2^{-1}$, compared with \$53.1 and \$64.0 when the nuclear lifecycle term is priced. The lifecycle inventory remains nonzero, and export credits still determine the negative accounting balances. These results apply to the specified carbon-accounting scenario; policy-specific eligibility and export-credit rules are not modeled.

\end{document}